\documentclass[reprint,amsmath,amssymb,pra,aps, nofootinbib]{revtex4-1}

\usepackage{amsmath, amssymb}
\usepackage{bbold}
\usepackage{bm}
\usepackage{tipa}
\usepackage{blindtext}
\usepackage[dvipsnames]{xcolor}
\definecolor{dcyan}{RGB}{0,100,100}
\usepackage{graphicx}
\usepackage{tabularx}
\usepackage{empheq}

\newcommand{\SLL}{\texttt{Landau cavity}}
\newcommand{\OT}{\texttt{Laughlin cavity}}

\begin{document}
\title{Aberrated optical cavities}
\author{Matt Jaffe$^1$}
\author{Lukas Palm$^1$}
\author{Claire Baum$^1$}
\author{Lavanya Taneja$^1$}
\author{Jonathan Simon$^1$}
\affiliation{$^1$The Department of Physics, The James Franck Institute, and The Pritzker School of Molecular Engineering, The University of Chicago, Chicago, IL}

\date{\today}

\begin{abstract}
        
        Optical cavities are an enabling technology of modern quantum science: from their essential role in the operation of lasers, to applications as fly-wheels in atomic clocks and interaction-enhancing components in quantum optics experiments, developing a quantitative understanding of the mode-shapes and energies of optical cavities has been crucial for the growth of the field. Nonetheless, the standard treatment using paraxial, quadratic optics fails to capture the influence of optical aberrations present in modern cavities with high finesse, small waist, and/or many degenerate modes. In this work, we compute the mode spectrum of optical resonators, allowing for both non-paraxial beam propagation and beyond-quadratic mirrors and lenses. Generalizing prior works~\cite{Laabs1999, visser2005spectrum, Klaassen2004, Klaassen2006, zeppenfeld2010calculating}, we develop a complete theory of resonator aberrations, including intracavity lenses, non-planar geometries, and arbitrary mirror forms. Harnessing these tools, we reconcile the near-absence of aberration in Ref.~\cite{schine2016synthetic} with the strongly evident aberrations in the seemingly similar cavity of Ref.~\cite{clark2020observation}. We further validate our approach by comparison to a family of non-planar lens cavities realized in the lab, finding good quantitative agreement. This work opens new prospects for cavities with smaller waists and more degenerate modes.
\end{abstract}

\maketitle

\section{Introduction}
Optical resonators have become an indispensable tool in optical and atomic physics. They are typically understood and utilized in the paraxial, quadratic limit, where the transverse mode structure and spectrum are derived~\cite{siegman86}. For common cavities, this results in the familiar Hermite-Gauss (HG) or Laguerre-Gauss (LG) families of eigenmodes, whose evenly-spaced resonance frequencies are set by the Gouy phase. Usually this description is entirely adequate, as deviations from this approximation are typically small when the resonator mode waist $w$ is much larger than the wavelength $\lambda$. Additionally, most applications use only the fundamental resonator mode.

As degenerate cavities have become more prevalent in quantum science experiments~\cite{kollar2015adjustable, schine2016synthetic,vaidya2018tunable,schine2019electromagnetic,clark2020observation}, the atomic physics community has begun to probe the limits of the aforementioned approximations by pursuing high-finesse, small waist resonators for their enhanced light-matter coupling~\cite{hunger2010fiber,tanji2011interaction,durak2014diffraction}. Degenerate cavities in particular are alarmingly sensitive to small deviations from the uniformly-spaced energy spectrum of the quadratic approximation. For the $n^{\text{th}}$ degenerate mode to overlap within a resonator linewidth, the spacing must be uniform to about one part in $n$ times the finesse. Optical resonators often have finesses in the range of $10^{3}$ to $10^{5}$, so achieving degeneracy requires extreme uniformity of the spectrum.

A number of attempts have been made to predict resonator spectra beyond the paraxial, quadratic limit. One might anticipate that finite-element or boundary-element approaches would provide quantitatively accurate results, but the fact that the resonators are many thousands of $\lambda$ across makes discretization a substantial computational challenge. Instead, the authors of Ref.~\cite{zeppenfeld2010calculating} make closed-form predictions for mode energies by expanding the mode functions of two mirror resonators in spheroidal coordinates. Refs.~\cite{visser2005spectrum,Klaassen2006} take a different approach, computing order-of-magnitude estimates of the impact of perturbations to the the paraxial resonator modes. Ref.~\cite{Laabs1999} analyzed nonparaxial eigenmodes of a half-symmetric two-mirror cavity using a perturbative expansion in a basis of HG modes. In this work, we pursue a more general treatment of aberrations using a novel perturbative expansion of the round-trip propagation operator in the basis of the paraxial quadratic eigenmodes. We find that this perturbative approach accurately captures observed mode-mixing \cite{schine2016synthetic, clark2020observation} arising from aberration terms near degeneracy. Additionally, we compare the model's predictions against measured spectra from new non-planar lens cavities, where we find quantitative agreement.

The structure of this paper is as follows: in Section~\ref{sec:base_formalism} we introduce the language that we will use to describe resonators. We then apply this formalism to describe the paraxial, quadratic case in Section~\ref{sec:paraxial_operator_calculation}. In Section~\ref{sec:perturbative_method} we extend the tools developed in the previous section by introducing a perturbative method for calculating aberrated spectra. We then validate our new tools by studying cubic astigmatism in a twisted resonator in Section~\ref{sec:cubic}, reconciling the previously opaque difference between two cavity designs showing either near-absence~\cite{schine2016synthetic} or strong presence~\cite{clark2020observation} of aberrations, understood only through a beyond-quadratic treatment. In Section~\ref{sec:lens_cav}, we compare computed and measured spectra of axis-symmetric twisted resonators containing intracavity lenses. Section~\ref{sec:summary} concludes. Technical aspects are presented in more detail in the appendices.

\section{The formalism} \label{sec:base_formalism}

We begin with an overview of the ray transfer (or ABCD) matrix approach for describing optical resonators. Consider a ray $\bm{v} = \begin{pmatrix} \bm{x} \\ \bm{s} \end{pmatrix} $ in the transverse plane, where $\bm{x}$ is the ray's position and $\bm{s}$ is the ray's slope ($\bm{x}$ and $\bm{s}$ can be taken to be either 1- or 2-dimensional). Eigenmodes of the cavity are represented by the eigenvectors $\bm{\mu}_{j}^{\pm}$ of the resonator's round-trip $ABCD$ matrix $M_{\text{rt}}$:
\begin{equation}
	M_{\text{rt}} \bm{\mu}_{j}^{\pm} = \nu_{j}^{\pm} \bm{\mu}_{j}
\end{equation}
\noindent
where $\nu_{j}^{\pm} = \exp(\pm i\chi_{j})$ is the corresponding complex eigenvalue with unity norm, and whose phase $\chi_{j}$ is the round trip Gouy phase, and $j$ indexes the transverse dimensions.

For two transverse dimensions, a mode $u_{qmn}$ with axial index $q$ and transverse mode indices $m,n$ has resonance frequency

\begin{equation} \label{eq:parax_quad_spectrum}
	\frac{\omega_{qmn}}{2\pi} = \left(q + m\frac{\chi_{1}}{2\pi} + n\frac{\chi_{2}}{2\pi} + \frac{\phi_{0}}{2\pi} \right)\times\text{FSR},
\end{equation}
\noindent
where FSR is the resonator's free spectral range. That is, the spectrum consists of evenly-spaced higher order modes, with spacings determined by the Gouy phases $\chi_{i}$. The overall phase $\phi_{0}$ accounts for mode-independent phase offsets, such as a reflection phase from a dielectric mirror. More details on the ray formulation of paraxial, quadratic resonators can be found in Ref.~\cite{siegman86}.

Since ABCD matrices cannot directly extend beyond the paraxial, quadratic limit, we will next describe an operator approach that reduces to the ABCD matrix approach in the quadratic limit~\cite{Stoler1981, Habraken2007}, but which we can then extend perturbatively. We begin with the Helmholtz equation $\nabla^2 \psi+n^2 k^2 \psi=0$, with wavenumber $k \equiv \frac{2\pi}{\lambda}$ for wavelength $\lambda$. This equation can be formally integrated along $z$ to yield the propagation operator of a field from plane $z=0$ to plane $z=L$ (assuming only forward-propagating waves) ~\cite{khan2016quantum}:

\begin{eqnarray} \label{eq:prop_operator}
	\psi(z=L)&=&U_{\text{prop}}(L)\psi(z=0)\\
	U_{\text{prop}}(L) &\equiv& \exp{\left(i L\sqrt{n^2 k^2-k_\perp^2}\right)}
\end{eqnarray}

\noindent
with $k\equiv\frac{2\pi}{\lambda}$, and $k_{\perp}\equiv -i\nabla_\perp$, and $\nabla_\perp$ is the gradient in the transverse plane.

For fields propagating nearly paraxially in a uniform medium ($n=n_0$ constant), $k_{\perp}\ll k$, so $U_{\text{prop}}(L) \approx e^{i k L n_0}e^{i L\frac{\nabla_\perp^2}{2kn_0}}$. That is, at lowest order, free-space propagation induces a phase shift which is quadratic in the light's transverse momentum $p_\perp\equiv-i \hbar\nabla_\perp$.

If the index of refraction $n$ varies in space (e.g., as we enter or leave a lens), we can, in the vicinity of the spatial variation of $n$, briefly drop the $\nabla_\perp$ entirely in Eqn.~\ref{eq:prop_operator} (as its influence builds up only over a Rayleigh range), yielding the forward-propagating solution:

\begin{eqnarray}\label{eq:lens_operator}
    \psi(z=L)&=&U_{\text{lens}}\psi(z=0)\\
     U_{\text{lens}}&=&e^{i k \int_0^L dz\,n(z)}=e^{i k L n_0}e^{i k (n_{1}-n_{0}) T(x,y)}
\end{eqnarray}
\noindent
where we have assumed that outside of the lens of thickness $T(x,y)$, the index is $n_{0}$, and within the lens the index is $n_{1}$. That is, at lowest order the impact of a lens on an optical field is to induce a spatially-varying phase shift. Similarly, the action of a mirror can be captured by a spatially varying phase shift $U_{\text{mirror}}=e^{i 2 k n_0 T(x,y)}$, where $T(x,y)$ is now the height of the curved surface of the mirror relative to a reference plane.

Further simplifying to near the center of such a mirror (or lens), we can treat the optic's form as quadratic in space, yielding:

\begin{equation} \label{eq:mirror_operator}
	U_{\text{mirror}}(R_{x}, R_{y}) \approx \exp{\left[-i kn_{0}\left(\frac{x^2}{R_x}+\frac{y^2}{R_y}\right)\right]},
\end{equation}
\noindent
where $R_x$ and $R_y$ are the radii of curvature of the mirror along two principal axes $x$ and $y$.

To understand resonators at lowest order, we thus concatenate free-space propagation and mirror/lens interface operators, to assemble a full cavity round-trip operator, and then compute the eigenmodes of this operator (which are the cavity modes). The challenge is that while $x$ and $y$ commute and $\partial_x$, $\partial_y$ commute, $x$ and $\partial_x$ do not commute: $[x,\partial_x]=-1$ (similarly for $y$ and $\partial_y$). We need to perform operator algebra in the exponentials to make further progress. We will find that at quadratic order, these exponential operators are \emph{equivalent} to ABCD matrices. But having derived the results from a full operator formalism, it will be apparent how to break the assumptions above and include the impact of higher-order (aberration) terms.

To simplify notation before proceeding, we write the slope in direction $i$ of a propagating ray as  $s_{i}\equiv-i\lambdabar\partial_i$, and position as $r\equiv \left(x,y\right)$. Now we have $[r_{i},s_{j}]=i\lambdabar\delta_{ij}$, the usual commutation relation of $x$ and $p$ in quantum mechanics, except with Planck's constant $\hbar$ replaced by the reduced wavelength $\lambdabar$ (that is, $\hbar \rightarrow \lambdabar \equiv 1/k = \frac{\lambda}{2\pi}$).

Note that in this operator formalism, $\left( \hat{x}, \hat{y}, \hat{s}_{x}, \hat{s}_{y} \right)$ are position and slope \emph{field operators}, as opposed to the classical position and slope of the ABCD formalism\footnote{The hats indicating operators have been omitted up to this point. Context will be sufficient to determine if, e.g., $x$ indicates an operator or a classical position}. We denote this collection of operators by $\hat{\nu} \equiv \begin{pmatrix}x&y&s_{x}&s_{y} \end{pmatrix}^{\intercal}$.

\section{The paraxial, quadratic resonator} \label{sec:paraxial_operator_calculation}
In the ray transfer matrix formalism, resonator characteristics are calculated from the eigenspectrum of the round-trip ABCD matrix $M_{\text{rt}}$. In the operator formalism, we will look for the eigenspectrum of a round trip unitary operator $U_{\text{rt}}$, composed of individual operations such as free propagation (Eqn.~\ref{eq:prop_operator}) and optical interfaces (e.g., Eqn.~\ref{eq:mirror_operator}) comprising a round trip through the resonator. We can decompose this round trip operator as follows:
\begin{equation} \label{eq:Urt}
	U_{\text{rt}}= \prod_{n=1}^{N} U_{n} = \prod_{n=1}^{N} e^{i\hat{\nu}^\intercal M_{n}\hat{\nu}}e^{i\theta_{n}(\hat{\nu})}.
\end{equation}
\noindent
For each of the $N$ pieces of optical evolution (e.g., free-propagation, reflection, refraction), $M_{n}$ encodes the action of the $n^{\text{th}}$ element on the field\footnote{Note that $M_{n}$ is not \emph{itself} an ABCD matrix. Our subsequent calculations demonstrate that it is the generator of the corresponding ABCD matrix under matrix exponentiation.}. $\hat{\nu}^\intercal M_{n}\hat{\nu}$ then represents a general quadratic operator comprised of the position and slope operators. We drop linear terms in $x$ ($\partial_x$) and $y$ ($\partial_y$), as these can be ``gauged away'' by re-centering (tilting) the cavity axis.

The $\theta_{n}$ contain the beyond-quadratic (``aberration") terms. In this section, we will characterize resonators without aberration terms (that is, in the paraxial, quadratic limit where $\theta_{n} = 0$), which we will re-introduce in Section~\ref{sec:perturbative_method}.

\subsection{Operator transformations} \label{subsec:the_operators}
The next step is to expand the (as yet unknown) eigen-functions in an (as yet undetermined) 2D harmonic oscillator basis $\psi=\sum_{mn}C_{mn}(a^\dagger)^m(b^\dagger)^n|\Omega\rangle$, where $|\Omega\rangle$ is the lowest mode of the cavity (TEM$_{00}$ for simple cavities) and $a^\dagger$, $b^\dagger$ create the two types of excitations (for example, $x$- and $y$- Hermite-Gauss) on top of that lowest mode.

We can now apply the round-trip operator to the eigenmode:
\begin{equation}
	U_{\text{rt}} |\psi\rangle = \prod_{\ell}\sum_{mn} e^{i\phi_{\ell}} \exp{\left(i\hat{\nu}_\alpha\hat{\nu}_\beta
			M_{\alpha\beta}^{\ell} \right)} C_{mn} (a^{\dagger})^{m} (b^{\dagger})^{n} |\Omega\rangle
\end{equation}
\noindent
where $U_{\text{rt}}$ in this Section denotes the quadratic round trip operator (i.e., all $\theta_{\ell} = 0$).

By inserting identities of the form $U_{\text{rt}}^\dagger U_{\text{rt}}$ between all of the operators in the eigenfunction, we arrive at:
\begin{equation}
	U_{\text{rt}}|\psi\rangle=\sum_{mn}C_{mn}(U_{\text{rt}} a^\dagger U_{\text{rt}}^\dagger)^m(U_{\text{rt}} b^\dagger U_{\text{rt}}^\dagger)^n U_{\text{rt}}|\Omega\rangle.
\end{equation}
\noindent
In words: if the $a^\dagger$, $b^\dagger$, $a$ and  $b$ operators transform in a simple way under $U_{\text{rt}}$ (we will find that they transform \emph{into each other}), we have made progress.

This is most directly addressed by following the approach of Eqn.~2.3.10b in Ref.~\cite{schumaker1986quantum}, where it is shown that $\exp{\left(i\hat{\nu}_\alpha\hat{\nu}_\beta M_{\alpha\beta}^m\right)}\hat{\xi}_\gamma \exp{\left(-i\hat{\nu}_\alpha\hat{\nu}_\beta M_{\alpha\beta}^m\right)}$, with $\hat{\xi}\equiv \begin{pmatrix}a&b&a^{\dagger}&b^{\dagger} \end{pmatrix}^{\intercal}$, is a Boguliubov transformation that can be computed in closed form. The central realization is that $\hat{\nu}$ (the position and momentum operators) may be written in terms of $\hat{\xi}$ (the raising and lower operators), via an as-yet undetermined transformation matrix $B$: $\hat{\nu}_k=B_{kl} \hat{\xi}_l$. Indeed, any such basis of operators is permissible so long as it satisfies the commutation relations: we will eventually want to choose a basis of raising and lowering operators that transform into themselves under $U_{\text{rt}}$, which is a more stringent constraint that we will address in Sec. ~\ref{sec:GettingB}.

Defining $N^{\ell}_{\mu\nu}\equiv M^{\ell}_{\alpha\beta}B_{\alpha\mu}B_{\beta\nu}$, we now need to compute $e^{i\hat{\xi}_\mu\hat{\xi}_\nu N_{\mu\nu}^{\ell}}{\hat{\xi}_\gamma} e^{-i\hat{\xi}_\mu\hat{\xi}_\nu N_{\mu\nu}^{\ell}}$. This is most simply achieved by employing Hadamard's lemma to the Baker-Campbell-Hausdorff identity: $e^A B e^{-A}=e^{\textrm{Ad}_A}B$, where $\textrm{Ad}_U V\equiv [U,V]$. Now noting that $[\hat{\xi}_\alpha,\hat{\xi}_\beta]=\left(i\sigma_y\otimes \mathbb{1}\right)_{\alpha\beta}\equiv G_{\alpha\beta}$ (where the identity operator operates on the a/b subspace and $i\sigma_y$ operates on the operator/adjoint subspace), we can write $[i \hat{\xi}_{\alpha}\hat{\xi}_{\beta} N_{\alpha\beta},\hat{\xi}_{\gamma}] = i N_{\alpha\beta} \left(\hat{\xi}_{\alpha} G_{\beta\gamma} + \hat{\xi}_{\beta} G_{\alpha\gamma}\right) = \left[ iG^\intercal(N^\intercal+N)\hat{\xi}\right]_\gamma = \left[2iG^\intercal N\hat{\xi}\right]_{\gamma}$, where we have used that $N_{\alpha\beta}$ is symmetric with respect to interchange of $\alpha$ and $\beta$ if $M$ is similarly symmetric.

We now arrive at:
\begin{equation}
	\exp{\left(i\hat{\nu}_\alpha\hat{\nu}_\beta M_{\alpha\beta}^{\ell}\right)}\hat{\xi}_\gamma \exp{\left(-i\hat{\nu}_\alpha\hat{\nu}_\beta M_{\alpha\beta}^{\ell}\right)} = \left[e^{2iG^\intercal N^{\ell}}\hat{\xi}\right]_{\gamma},
\end{equation}
\noindent
which is the central result Eqn.~2.3.10b of~\cite{schumaker1986quantum}, and shows that the $\hat{\xi}$ transform into one another under paraxial/quadratic propagation according to a Boguliubov transformation.

A nearly-identical calculation can be performed to see how the position and slope operators $\hat{\nu}$ transform. We take the commutation relation $[\hat{\nu}_\alpha,\hat{\nu}_\beta] = i \lambdabar G_{\alpha\beta}$, and can use $M_{\alpha\beta}$ directly, without having to introduce $N_{\alpha\beta}$. The result is that for a transformation encoded in $M$, $\hat{\nu}$ transforms according to
\begin{align}
    U \hat{\nu}_{\gamma} U^{-1} &\equiv \exp{\left(i\hat{\nu}_\alpha\hat{\nu}_\beta M_{\alpha\beta} \right)} \hat{\nu}_\gamma \exp{\left(-i\hat{\nu}_\alpha\hat{\nu}_\beta M_{\alpha\beta} \right)} \nonumber \\
    &= \left[e^{2\lambdabar G M}\hat{\nu}\right]_{\gamma}.
\end{align}
For the 1D case of a mirror, we have that $ M = \begin{pmatrix} -k/R & 0 \\ 0 & 0 \end{pmatrix}$, so after exponentiation, $\hat{\nu}\rightarrow\begin{pmatrix} 1 & 0 \\ -2/R & 1 \end{pmatrix}\hat{\nu}$, which is the ABCD matrix evolution for reflection off of a curved mirror~\cite{siegman86}. Similarly, the evolution of $\hat{\nu}$ under free-propagation $M=\begin{pmatrix} 0 & 0 \\ 0 & -kL/2 \end{pmatrix}$ obeys its ABCD matrix evolution $\hat{\nu} \rightarrow\begin{pmatrix} 1 & L \\ 0 & 1 \end{pmatrix}\hat{\nu}$. We see here that $M$ generates the familiar ABCD matrix of a ray transformation~\cite{Wolf2004}.

In summary, we have shown that within the paraxial/quadratic approximations, the position and slope field operators obey the same ABCD matrix evolution as the classical ray position and slope~\cite{siegman86}.

\subsection{The Gouy phases}

Having ascertained how $\hat{\nu}$ and $\hat{\xi}$ evolve under paraxial/quadratic propagation, we can now ask how cavity eigenmodes $|\psi_{mn}\rangle\equiv\frac{1}{\sqrt{m!n!}}a^{\dagger m}b^{\dagger n}|\Omega\rangle$ evolve under $U_{\text{rt}}$. The central realization is that we must derive the eigen-operators $a^\dagger$ and $b^\dagger$ of $U_{\text{rt}}$, satisfying $U_{\text{rt}}c^\dagger U_{\text{rt}}^\dagger=\lambda c^\dagger$, for $c\in [a,b]$. Since $\hat{\xi}=B^{-1}\hat{\nu}$, the eigenvalues for the round-trip evolution of $\hat{\nu}$ are the same as those of $\hat{\xi}$, and we know $\hat{\nu}$ obeys $U_{\text{rt}}\hat{\nu} U_{\text{rt}}^{-1}=\prod_{\ell}e^{2\lambdabar G M^{\ell}}\hat{\nu}$ from the preceding subsection, so the eigenvalues of the round trip ABCD matrix are the eigenvalues of the (still unknown) $a^\dagger$ and $b^\dagger$ operators. It is well known that these eigenvalues come in complex-conjugate pairs for stable optical resonators $\lambda=e^{\pm i\phi_{a,b}}$ so we write the eigenvalues associated with the operators $a^\dagger$ and $b^\dagger$ as $e^{i\phi_{a,b}}$.

It is now apparent that the round-trip evolution of $|\psi_{mn}\rangle$ is given by
\begin{align}
	U_{\text{rt}}|\psi_{mn}\rangle  &= \frac{1}{\sqrt{m!n!}}a^{\dagger m}b^{\dagger n}e^{i(m\phi_a+n\phi_b)}U_{\text{rt}}|\Omega\rangle \nonumber \\
							&= e^{i(m\phi_a+n\phi_b + \phi_{0})}|\psi_{mn}\rangle
\end{align}
\noindent
where $e^{i \phi_{0}}$ is the eigenvalue of the lowest-order mode $|\Omega\rangle$ under the round-trip operator $U_{\text{rt}}$.

While we have not yet computed the eigen-operators $a^\dagger$, $b^\dagger$, nor the lowest order eigenmode $|\Omega\rangle$ and its eigenvalue $e^{i\phi_0}$, we can still say $\phi_{mn}=m\phi_a+n\phi_b+\phi_0$; that is, $\phi_a$ and $\phi_b$ are the round-trip ``Gouy phases'' (including a geometric phase contribution from round-trip axis rotation). $\phi_0$ is a phase that we can compensate for by nanoscopically modifying the resonator length, and is indeed very difficult to measure for the same reason.

\subsection{Computing $a,a^\dagger,b,b^\dagger$ operators invariant under $U_{\text{rt}}$} \label{sec:GettingB}

We managed to prove the results of the previous section without computing $B$ or $|\Omega\rangle$. This is because in the paraxial, quadratic approximation, $x$ and $s$ evolve linearly into one another, so we can entirely avoid asking about anything that depends upon the zero-point motion, non-commutativity of $x$ and $s$, or mode functions. It is perhaps surprising, and definitely quite useful, that we could extract the spectrum of the system without computing the modes; this is because the mode-spectrum is just given by the classical harmonic oscillator frequencies. On the other hand, once we introduce higher order terms into the problem, it will be impossible to avoid the impact of the mode-functions on  the spectrum. As such, we will need to now compute $B$ and $|\Omega\rangle$.

We would like to find an operator $\hat{A}=\chi^\intercal\hat{\nu}$ (i.e., a sum over the position/slope operators $\hat{\nu}$ with some coefficient list $\chi$) such that $U_{\text{rt}}\hat{A}U_{\text{rt}}^{-1}=\lambda\hat{A}$; this will ultimately become one of our two lowering operators $a$ and $b$. Remembering that the position and slope operators $x,y$ and $s_x,s_y$, combined as $\hat{\nu}$, transform according to $U_{\text{rt}}\hat{\nu} U_{\text{rt}}^{-1} = M_{\text{rt}} \hat{\nu}$ (for round trip ABCD matrix $M_{\text{rt}}$), and $\chi^\intercal$, a constant row-vector, is invariant under $U_{\text{rt}}$, we have that $U_{\text{rt}}\hat{A}U_{\text{rt}}^{-1}=\chi^\intercal  M_{\text{rt}}\hat{\nu}=\lambda\chi^\intercal \hat{\nu}$, meaning that $\chi^\intercal$ is a left-eigenvector of $M_{\text{rt}}$, or equivalently, a row of the inverse of the matrix of right-eigenvectors. In fact, as shown in Eqn.~2.60 of Ref.~\cite{habraken2010light}, the left and right eigenvectors of $M_{\text{rt}}$ have a simpler relationship for physical ABCD matrices $M_{\text{rt}}$ which must obey the defining property $M_{\text{rt}}^\intercal G M_{\text{rt}}=G$ of the symplectic group $\text{Sp}(2n, \mathbb{R})$ (where $G=\begin{pmatrix} 0 & \mathbb{1} \\-\mathbb{1} & 0\end{pmatrix}$ is the matrix from Section~\ref{subsec:the_operators}). It is then straightforward to show that if $\mu_{\text{rt}}$ is a right-eigenvector of $M_{\text{\text{rt}}}$, then $\chi^\intercal\equiv\mu_{\text{rt}}^\intercal G$ is a left-eigenvector with the inverse eigenvalue. In other words, up to a normalization constant, the raising (or lowering, as yet unknown) operator is $\hat{A}=\mu_{\text{rt}}^\intercal G\hat{\nu}$. It bears mention that we could have just worked with left-eigenvectors, but chose not to by convention.

We prove the preceding statement by starting with $M_{\text{rt}}\mu_{\text{rt}}=e^{i\phi}\mu_{\text{rt}}$. Taking the transpose of this equation yields $\mu_{\text{rt}}^\intercal M_{\text{rt}}^\intercal=e^{i\phi}\mu_{\text{rt}}^\intercal$; replacing $M_{\text{rt}}^\intercal=-G M_{\text{rt}}^{-1}G$ and noting $G^2=-1$ yields $\mu_{\text{rt}}^\intercal G M_{\text{rt}}^{-1}=e^{i\phi} \mu_{\text{rt}}^\intercal G$. Finally, we right-multiply by $M_{\text{rt}}$ and divide by $e^{i\phi}$ to arrive at: $e^{-i\phi}\mu_{\text{rt}}^\intercal G=\mu_{\text{rt}}^\intercal G M_{\text{rt}}$. That is, $\mu_{\text{rt}}^\intercal G$ is a right-eigenvector of $M_{\text{rt}}$ with eigenvalue $e^{-i\phi}$.

The normalization of $\hat{A}$ requires $[\hat{A}^\dagger,\hat{A}]=1$; employing full index notation for $\hat{A}=\mu^{\text{rt}}_i G_{ij}\hat{\nu}_{j}$, then $[\hat{A}^\dagger,\hat{A}]=-G_{ij}\mu^{\text{rt}*}_j G_{kl}\mu^{\text{rt}}_k[\hat{\nu}_{i},\hat{\nu}_{l}]$, and noting that $[\hat{\nu}_{i},\hat{\nu}_{l}]=i\lambdabar G_{il}$, then $[\hat{A}^\dagger,\hat{A}]=-i\lambdabar G_{ij}\mu^{\text{rt}*}_j G_{kl}\mu^{\text{rt}}_k G_{il}=-i \lambdabar \mu_{\text{rt}}^\dagger G \mu_{\text{rt}}=1$, means that the proper normalization requirement is $\mu_{\text{rt}}^\dagger G\mu_{\text{rt}}=i\frac{2\pi}{\lambda}$. Note that half of the eigenvectors of $M_{\text{rt}}$ will produce a normalization of $-i\frac{2\pi}{\lambda}$: these are instead the raising operators, since the eigenvectors come in complex-conjugate pairs, and indeed this is how we choose which two of the four eigenvectors to use to generate the two lowering operators.

We now have all the necessary information to generate $B$. Recall that $B$ allows us to write the position and slope operators contained in $\hat{\nu}$ in terms of the mode raising and lowering operators contained in $\hat{\xi}$: $\hat{\nu}=B\hat{\xi}$.  The rows of $B^{-1}$ are thus (with $\mu_{\text{rt}}$ now properly normalized):
\begin{equation} \label{eqn:raisingoperators}
	\frac{\mu_{\text{rt}}^\intercal G}{\sqrt{-i \lambdabar \mu_{\text{rt}}^\dagger G \mu_{\text{rt}}}}
\end{equation}

\subsection{Computing the lowest transverse mode of the cavity} \label{sec:GettingOmega}
We will make an ansatz that the lowest mode of the cavity is a (properly normalized) Gaussian $|\Omega\rangle=\frac{|W|}{2\pi}e^{-x_{\alpha} x_{\beta} W_{\alpha\beta}/2}$. For this to be the lowest mode, it must be annihilated by both $a$ and $b$ -- though we will not prove it, this is a sufficient condition to be a cavity eigenmode as well.

As above, we write $a$ and $b$ in terms the right- eigenvectors of $M_{\text{rt}}$, $\mu_{\text{rt}}^{a,b}$, according to $a/b=\frac{\mu_{\text{rt}}^{a/b\intercal} G}{\sqrt{-i \lambdabar \mu_{\text{rt}}^{a/b\dagger} G \mu_{\text{rt}}^{a/b}}}=\rho_{a/b}^\intercal\hat{\mu}$, which define $\rho_{a/b}$. The requirements $a|\Omega\rangle=b|\Omega\rangle=0$ become (imposing that $W$ is symmetric): $W=iU^{-1}V/\lambdabar$, where $\left( U V\right)=\begin{pmatrix} \rho_a^\intercal\\ \rho_b^\intercal\end{pmatrix}$ defines $U$ and $V$ (which could equally well be computed using the un-normalized $\mu_{\text{rt}}^{a/b\dagger} G$).

The definitions of $a$, $b$ and $|\Omega\rangle$ all depend upon the chosen reference plane (the plane that we return to after a cavity round trip) -- within the quadratic/paraxial approximation different reference planes are related by a fractional Fourier transform, and in general changing reference planes corresponds to the Floquet micromotion~\cite{sommer2016engineering}. By contrast, the Gouy phases, and hence the transverse mode spacings, do not depend upon the chosen reference plane.

In summary, we have described the paraxial, quadratic resonator using an operator formalism. We solved for the Gouy phases $\phi_{a,b}$ (and thus the spectrum), as well as the lowest transverse mode $|\Omega \rangle$. Importantly, we calculated the raising and lowering operators $\hat{\xi}\equiv \begin{pmatrix}a&b&a^{\dagger}&b^{\dagger} \end{pmatrix}^{\intercal}$ in terms of the position and slope operators $\hat{\nu} \equiv \begin{pmatrix}x&y&s_{x}&s_{y} \end{pmatrix}^{\intercal}$ and the eigenvectors $\mu_{\text{rt}}$ of the cavity round trip ABCD matrix $M$. In the next section, we extend this formalism to describe aberrations.

\section{Perturbative spectrum computation} \label{sec:perturbative_method}
Now that we have fully characterized the quadratic resonator within our operator framework, we can re-introduce resonator aberrations, $\theta_{n} \neq 0$ in Eqn.~\ref{eq:Urt}. To achieve this, we will decompose the round-trip operator for the fully aberrated (non-paraxial, non-quadratic) evolution in the basis of the quadratic eigen-modes. Except in very special cases, the higher-order terms will only weakly perturb the modes of the cavity, so this basis will be a extremely efficient choice.

In the non-paraxial/non-quadratic limit, the resonator eigenmodes are no longer uniformly spaced in energy, and indeed, a pair of raising operators that are linear in $x$ and $s$ will no longer generate the modes by repeated application to the lowest mode. In fact, the sense  of a ``lowest mode'' will itself break down due to the Floquet nature of the system~\cite{sommer2016engineering} and higher-order terms mixing the modes. We will only be able to identify the ``lowest mode'' by adiabatically connecting it to $|\Omega\rangle$ away from degeneracy points where the modes mix.

\subsection{Perturbative expansion of the aberrations}
Let us consider the evolution of a state decomposed into the basis of quadratic eigenstates in the $k^{\text{th}}$ reference plane (after the first $k-1$ propagation terms around the cavity, as parameterized in $U_{\text{rt}}$ above): $|\psi\rangle_k=\sum_{mn}C_{mn}^k(a_k^\dagger)^m(b_k^\dagger)^n|\Omega_k\rangle$. Here $|\Omega_k\rangle\equiv\prod_{n=1}^k e^{i\hat{\nu}^\intercal M_n\hat{\nu}}|\Omega\rangle$ is the maximally localized-quadratic eigenstate quadratically propagated to the $k^{\text{th}}$ reference plane, and:

\begin{equation} \label{eqn:definingplaneraisinglowering}
	\{a_k^\dagger,b_k^\dagger\}=\left(\prod_{n=1}^k e^{i\hat{\nu}^\intercal M_n\hat{\nu}}\right)\{a^\dagger,b^\dagger\}\left(\prod_{n=1}^k e^{-i\hat{\nu}^\intercal M_n\hat{\nu}}\right)
\end{equation}
\noindent
are the quadratic raising operators quadratically propagated to the $k^{\text{th}}$ reference plane; The $C_{mn}^k$ then, are the coefficients of the wavefunction in the $k^{\text{th}}$ reference plane.

We now propagate to the $(k+1)^{\mathrm{st}}$ reference plane, to connect $C_{mn}^k$ to $C_{mn}^{k+1}$:
\begin{align*}
	|\psi\rangle_{k+1} 	& \equiv U_{k}|\psi\rangle_k\\
					 	& = \sum_{mn} e^{i\hat{\nu}^\intercal M_k\hat{\nu}}e^{i\theta_k(\hat{\nu})} C_{mn}^k(a_k^\dagger)^m(b_k^\dagger)^n|\Omega_k\rangle
\end{align*}

To proceed, we note that the $\theta_{n}$ are generically quite complex to compute, so for now we will assume that we know them in terms of $x,y$ and $p_x,p_y$, and leave their general computation for later sections and appendices.

Because the $\theta_k(\hat{\nu})$ can be written in terms of $x,y,p_x,p_y$, they can equivalently be written in terms of the raising and lowering operators in the $k^{\text{th}}$ reference plane, by utilizing $\hat{\nu}=B_k\hat{\xi}_k$. We then explicitly write out $\hat{\xi}_k$ in the excitation number basis, truncating at a finite excitation number, and can thus write $e^{i \theta_{k}}$ as a matrix-exponential of $i\theta_{k}$ in this same (truncated) number basis. The truncation is acceptable here because we have assumed that the number-basis of the quadratic eigenstates is nearly correct, and the $\theta_{k}$ only generate a weak perturbation. In practice, this results in an expression of the form $e^{i\theta_{k}}(a_k^\dagger)^m(b_k^\dagger)^n\approx\sum_{pq}D^k_{mnpq}(a_k^\dagger)^p(b_k^\dagger)^q$, where the $D^k_{mnpq}$ are the matrix elements of the (potentially highly-nonlinear) $e^{i\theta_k}$ in the number-basis of the $k^{\text{th}}$ reference plane.

We now have:
\begin{equation*}
	U_{k}|\psi\rangle_{k} = \sum_{mnpq} e^{i\hat{\nu}^\intercal M_k\hat{\nu}}e^{i\theta_k(\hat{\nu})} D^k_{mnpq}C_{mn}^k(a_k^\dagger)^p(b_k^\dagger)^q|\Omega_k\rangle
\end{equation*}
and defining $C^{k+1}_{mn}=\sum_{ij}D_{ijmn}C_{ij}^k$, we have $U_{k}|\psi\rangle_k=\sum_{pq} C_{pq}^{k+1} e^{i\hat{\nu}^\intercal M_k\hat{\nu}}(a_k^\dagger)^p(b_k^\dagger)^q|\Omega_k\rangle$. Multiplying through by the identity $1=V_k V_k^\dagger$, with $V_k\equiv e^{i\hat{\nu}^\intercal M_k\hat{\nu}}$, we find $U_{k}|\psi\rangle_k=\sum_{pq} C_{pq}^{k+1} (V_k a_k^\dagger V_k^\dagger)^p(V_k b_k^\dagger V_k^\dagger)^q V_k|\Omega_k\rangle$. We can now identify $|\Omega_{k+1}\rangle$, $a^\dagger_{k+1}$,$b^\dagger_{k+1}$, and arrive at the expected final result: 

\begin{equation} \label{eqn:fullaberrationresult}
	U_{k}|\psi\rangle_k=\sum_{pq} C_{pq}^{k+1} (a_{k+1}^\dagger)^p(b_{k+1}^\dagger)^q|\Omega_{k+1}\rangle
\end{equation}

We may understand the $D^k_{mnpq}$ as matrix representations of operators $D_k$ that take a wavefunction from mode $(m,n)$ (the state $(a^\dagger)^m(b^\dagger)^n|\Omega\rangle$) to mode $(p,q)$. The last remaining step is to project back from our final raising/lowering operator basis $a_N^\dagger,a_N,b_N^\dagger,b_N$ to our initial basis $a_0^\dagger\equiv a^\dagger,a_0\equiv a_,b_0^\dagger\equiv b^\dagger,b_0\equiv b$. As the $N^{\text{th}}$ reference plane is the same as the original reference plane, and we have chosen our raising/lowering operators to be eigen-operators of $U_{\text{rt}}$, the projection just extracts the Gouy phases, through the matrix $Q^{\text{rt}}_{mnpq}=\delta_{mp}\delta_{nq}e^{i(\phi_a m+\phi_b n)}$.

In total, then $U_{\text{rt}}=Q_{\text{rt}}\times\prod_{k=0}^N D_k$ is the round-trip operator including all aberrations, in the basis of the instantaneous quadratic eigenstates. Finding the eigenvalues/vectors of this matrix provides the full resonator spectrum.\\

\subsection{Perturbation forms}
The last challenge that remains is to specify the form of the perturbations: \emph{how should we explicitly specify the $\theta_{k}(\hat{\nu})$ associated with a given perturbation}?

In the case of non-paraxial propagation, the answer is straightforward: we simply expand $U_{\text{prop}}(L) \equiv \exp{\left(i L\sqrt{n^2 k^2-k_\perp^2}\right)}$ to higher order in $s\propto k_\perp$ arriving at $\theta_{\text{nonparax}}^{[4]} = -\frac{kL}{8}s^{4}$, at quartic order.

For higher-order corrections to the behaviors of lenses and mirrors, the situation is substantially complicated by the fact that the \emph{shape} of the surface produces aberrations at the same order as the non-paraxial propagation of the beam. Put another way: describing the optic as just a position-dependent phase plate omits momentum ($\partial_{x}$)-dependent effects that are important at the same order. We can \emph{formally} write the operator describing light propagating through an optical interface in terms of a $z$-ordered product~\cite{khan2018aberrations}:

\begin{eqnarray}\label{eq:full_lens}
    U_{\text{lens}} &= &\mathcal{Z} \left\{\exp\left[i \int_{0}^{L} dz\,\sqrt{k^{2} n^2(x,y,z)+\nabla_\perp^2}\right]\right\}
\end{eqnarray}

Further progress from here is challenging, and is the topic of active research in application of Baker-Campbell-Hausdorff identities, Magnus expansions, and even analogies to relativistic quantum mechanics~\cite{khan2016quantum, khan2018aberrations, Dragt1986, Navarro-Saad1986, GrpThrtcI1986, GrpThrtcII1986, GrpThrtcIII1987}.

Our central proposition is that Hamiltonian optics and ray-tracing together accurately provide the perturbation polynomials at the next non-vanishing order. These perturbation forms and their derivations can be found in Appendix~\ref{appendix:pert_polys}. The proposed approach is validated by comparison to the experiments of Sections ~\ref{sec:cubic} and ~\ref{sec:lens_cav}. We find that this view provides a consistent physical picture and accurately describes experimental data.

\section{Non-planar curved-mirror cavities: the role of cubic astigmatism} \label{sec:cubic}
To demonstrate the utility of this perturbative approach, we will calculate the aberrations of a resonator whose degeneracy is broken, at lowest order, by cubic astigmatism. The resonators summarized in Table~\ref{table:cav_specs}, developed to explore lowest Landau level physics and Laughlin states of photon pairs, were specifically designed to suppress the impact of \emph{quadratic} astigmatism~\cite{schine2016synthetic,schine2019electromagnetic}. In each case, the Landau level is formed by a set of degenerate orbital angular momentum (OAM) modes. The first of the two cavities, with a larger waist, exhibited no observable avoided crossing near degeneracy~\cite{schine2016synthetic}; the second, with a $\sim 2 \times$ reduced mode waist size, presented clear avoided crossings as degeneracy was approached~\cite{clark2020observation}. In short, these cavities provide a clear and simple testbed for beyond-quadratic resonator aberrations.

\begin{figure*} 
	\centering
	\includegraphics[width=\textwidth]{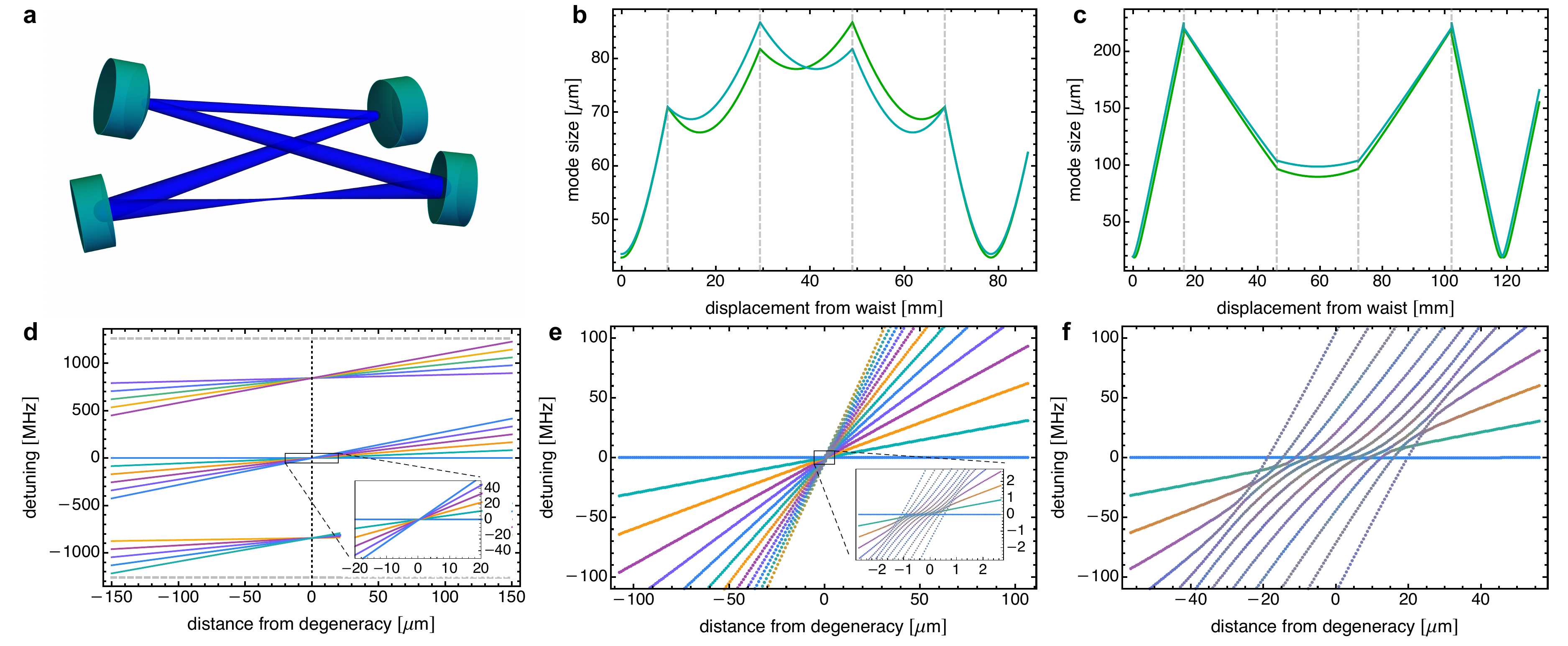}
	\caption{Two non-planar cavities are described: {\SLL}~\cite{schine2016synthetic} and {\OT} \cite{clark2020observation}. (a) Non-planar cavity geometry. {\OT} is shown, but {\SLL} is nearly identical (differences are outlined in Table~\ref{table:cav_specs}). (b, c) Mode size over a cavity round trip for {\SLL} and {\OT}, respectively for $\lambda = 780$ nm. Mirror positions are indicated by the dashed vertical lines. Colors indicate the two semi-axes. Small discontinuities are due to a change of semi-axis basis arising from the astigmatism plus rotation after reflection into a new plane. (d) Paraxially-expected spectrum, with target degeneracy inset. (e,f) Aberrated theory spectra for {\SLL} and {\OT}, respectively. Each point's color is a blend of the color scheme in (d), with weightings given by the paraxial eigenmode contributions to the point's corresponding eigenvector. Level repulsion and mode mixing due to cubic astigmatism can be seen, which is much stronger in {\OT}. Refer to the main text for further discussion.}
	\label{fig1}
\end{figure*}

\begin{table}
	\caption{Overview of the cavities under comparison. Full details can be found in Ref.~\cite{SchineThesis}.}
	\begin{center}
		\begin{tabular}{| l | c | c|}
			\hline
			&   \SLL  &   \OT  \\
			\hline
			waist     		&     	43 $\mu$m 			& 			19 $\mu$m      	\\
			Mirror ROCs [mm]		& $\{25,50,50,25\}$ &  $\{25,-75,-75,25\}$	\\
			round-trip length  & 		79 mm 		& 			119 mm		\\ 
			\hline
		\end{tabular}
	\end{center}
	\label{table:cav_specs}
\end{table}

\begin{figure*} 
	\centering
	\includegraphics[width=\textwidth]{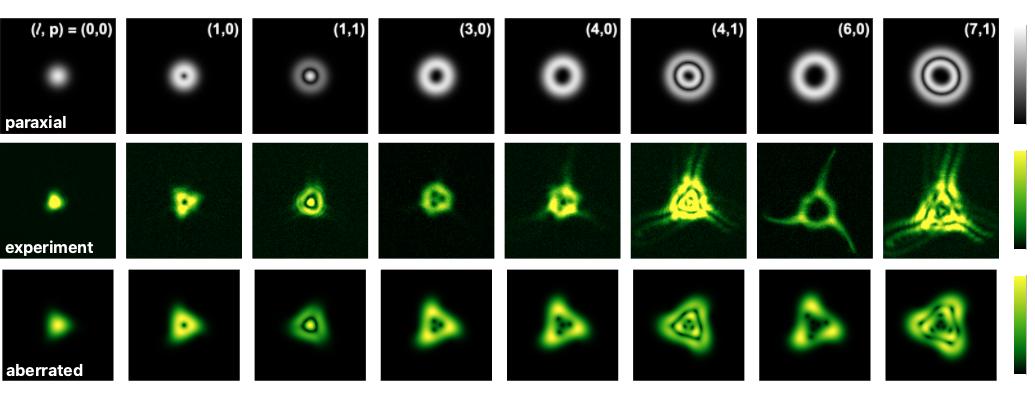}
	\caption{Aberrated mode profiles. Mode profiles are shown for {\SLL} near the degeneracy point shown in Fig.~\ref{fig1}d. Profiles are shown for (top row) paraxial expectation (near-Laguerre-Gauss modes, with mode indices indicated), (middle row) experimentally measured profiles, and (bottom row) aberrated calculation. Color bars at right scale between zero and peak intensity of each image.}
	\label{fig2}
\end{figure*}

The cavities in questions use a non-planar twist to generate a synthetic magnetic field for light~\cite{schine2016synthetic,sommer2016engineering,schine2019electromagnetic}. The non-planar twist necessitated off-axis incidence on curved mirrors, and thus exhibited quadratic astigmatism due to  the different effective radii of curvature for the sagittal and tangential axes~\cite{siegman86}. In the plane of the mirror, this astigmatism can be represented by an operator polynomial in position $U^{[2]}_{\text{astig}}\propto x^{2} - y^{2} = r^2 (e^{i 2\theta}+e^{-i 2\theta})$; from the second expression, it is clear that quadratic astigmatism couples every \emph{second} OAM mode. Since a planar Landau level (in the symmetric gauge) consists of every OAM mode (without radial nodes), it is clear that its degeneracy will be destroyed by such quadratic astigmatism.

More formally in the operator picture, we have the relation $\hat{\nu} = B \hat{\xi}$, where $\hat{\nu} = \begin{pmatrix}x&y&s_{x}&s_{y} \end{pmatrix}^{\intercal}$ and $\xi = \begin{pmatrix}a&b&a^{\dagger}&b^{\dagger} \end{pmatrix}^{\intercal}$. Quadratic astigmatism can then be seen to contain terms like $aa, \, a^{\dagger} a^{\dagger}$, when using $B$ to write quadratic astigmatism in the excitation number basis. $a^{\dagger}a^{\dagger}$ couples states differing by two quanta along the $a$ axis. If modes coupled in this way approach degeneracy, the coupling becomes resonant, leading to mode-mixing and unstable/lossy cavity modes.

To avoid this destabilization, in Ref.~\cite{schine2016synthetic}, it was found that by ensuring that the twist generated a Gouy phase of $2\pi/3$, a conical Landau level could be realized, consisting of only every third OAM mode $l=0,3,6,...$, thereby suppressing the impact of quadratic astigmatism.

Unfortunately, this simply pushed the problem to slightly higher order: Non-normal incidence on a spherical surface \emph{also} introduces cubic astigmatic (see Appendix~\ref{sec:non_normal_aoi} for the form of this perturbation). We now compute the effects of cubic astigmatism on the resonator spectrum, distinguishing between the cavity of Ref.~\cite{clark2020observation} (we will call this cavity ``{\OT}") where these effects were apparent, and the seemingly-similar cavity of Ref.~\cite{schine2016synthetic} (``{\SLL}")\footnote{As they were used to demonstrate a Laughlin state of photons and Landau levels for photons, respectively}, where they were not.

The basic resonator configuration is shown in Fig.~\ref{fig1}a. Four mirrors are arranged in a tetrahedral configuration, providing the non-planarity. Specifications for each cavity are shown in Table~\ref{table:cav_specs}. The second generation cavity {\OT} was designed for a smaller waist (to allow for Rydberg-mediated interactions between the photons), and we will see that this dramatically increases the effect of aberrations.

The mode sizes for {\SLL} and {\OT} over a cavity round trip are shown in Figs.~\ref{fig1}b,c, respectively. The expected spectrum calculated under the paraxial, quadratic assumptions is shown in Fig.~\ref{fig1}d. Only the modes of interest, angular momentum modes with Laguerre-Gauss indices $(\ell,p) = (\ell, 0)$ are shown\footnote{Higher Landau level modes, with $p>0$, are discussed in Ref.~\cite{schine2016synthetic}}. As the Gouy phase varies with mirror spacing, a degeneracy is approached when the mirror spacing sets the total round trip Gouy phase to be $\frac{2\pi}{3}$. The expanded inset in Fig.~\ref{fig1}d shows the expected degeneracy of angular momentum modes $\ell = 0,3,6,...$.

Fig.~\ref{fig1}e,f shows the calculated perturbed spectra of {\SLL} and {\OT}, respectively. The cubic aberration is evident as the level repulsion around the expected degeneracy. Incomplete level repulsion of the highest-order modes shown is a finite basis effect (edge basis states do not have higher levels to couple to). In reality, high-order modes also see increasing loss, as larger modes run off the edge of the mirror, or encounter mirror imperfections within their larger surface area. For strong mixing, even ``low-order" modes become lossy, as they acquire a significant contribution of high-order unperturbed modes. In fact, this mixing was strong enough to destabilize even the lowest-order mode in {\OT} as degeneracy was reached.

Apparently, this modest reduction in waist size comes with a dramatic increase in the cubic aberration. This can also be seen in the round trip mode-size plots (Figs.~\ref{fig1}a,b) as the extra ``work'' done by each curved mirror surface in the aberrated geometry. The zoomed inset of Fig.~\ref{fig1}e shows a similar level structure to that of Fig.~\ref{fig1}f, though with much weaker mixing. Throughout Fig.~\ref{fig1}, we include only the cubic perturbation (i.e., we ignore quartic and higher non-paraxial propagation and spherical aberration terms). Our perturbative approach is limited by commutator ambiguities~\cite{Wolf2004} when combining terms of different orders, but we can ignore higher-order terms for these cavities, which are dominated by resonant cubic terms.

Our calculation method enables construction of the mode profiles (eigenvectors), in addition to the mode energies (eigenvalues) from $U_{\text{rt}}$. For example, the mixed-modes of {\SLL} near the degeneracy point in Fig.~\ref{fig1}d are shown in Fig.~\ref{fig2}. The resonant cubic astigmatism leads to a clear three-fold symmetry, as modes separated by 3 OAM quanta are coupled by cubic terms.

While the resemblance is clear, the experimental modes show a more dramatic deviation from the paraxial expectation than even the aberrated predictions. This could be due to further effects, such as (a) mixing that is strong enough to be non-perturbative, (b) interaction between cubic terms and higher-order terms (e.g., resonant 6th-order astigmatism [as indicated by the six-fold symmetry in some modes], non-paraxial propagation, spherical aberration), or (c) mode-dependent loss. High-order modes are clearly involved, as seen by the long tails extending out to large radii in the last few columns of Fig.~\ref{fig2}.

\section{Non-planar lens cavities: the role of axisymmetric aberration}\label{sec:lens_cav}

Motivated by the goals of the cavities in Sec.~\ref{sec:cubic}, we propose and build a twisted cavity consisting of flat mirrors and two intra-cavity lenses. Flat mirrors allow for the non-planar twist without introducing astigmatism via non-normal incidence, while the on-axis intracavity lenses provide the transverse confinement necessary for a stable cavity. This arrangement enables a non-astigmatic cavity without relying on hard-to-manufacture elliptical or off-axis parabolic mirrors\footnote{The surface roughness and anti-reflection coating of the lenses must support the required finesse.}.

	\begin{figure*} 
		\centering
		\includegraphics[width=\textwidth]{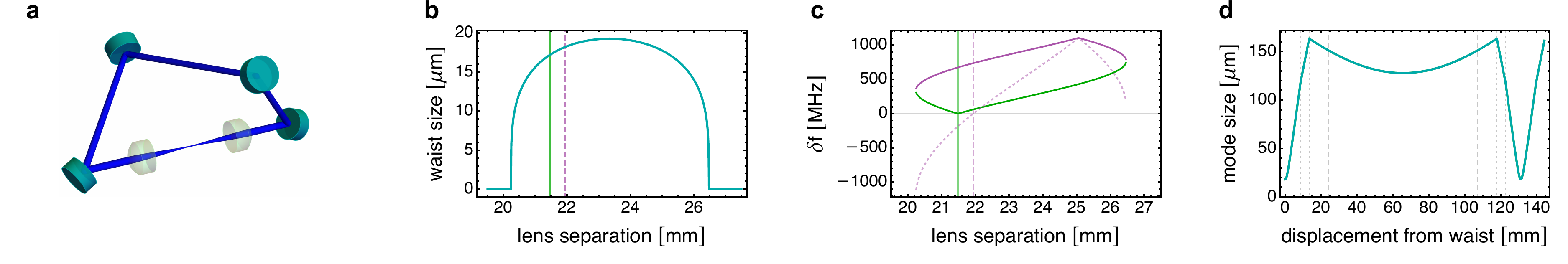}
		\caption{Twisted lens cavity. (a) Schematic of the cavity, showing the two intra-cavity lenses and the non-planar geometry. (b) Waist size over the stability region of the cavity as a function of intra-cavity lens spacing. (c) Transverse mode spacing along the two axes. There are degeneracy points at $s=1$ (green) and $s=3$ (purple). For $s=3$, $s$ times the transverse mode spacing is shown as a dashed line to illustrate its zero-crossing. Vertical lines indicate the degeneracy points of their respective color, and are reproduced in (b). (d) Mode size over a cavity round trip at $s=3$ degeneracy. Mirrors are indicate by dashed vertical lines, and front and back lens surfaces by dotted lines. For (b-d), $\lambda = 780$ nm.}
		\label{fig3}
	\end{figure*}

A schematic of the cavity layout can be seen in Fig.~\ref{fig3}a.
The cavity waist and transverse mode splittings as a function of lens separation are shown in Figs.~\ref{fig3}b,c, respectively. The cavity is designed to have one degeneracy point on each axis (i.e., where that axis' Gouy phase equals $\frac{2\pi}{s}$ for integer $s$), at $s=1$ and $s=3$. These points can be seen in Fig.~\ref{fig3}c, shown in green and purple, respectively. At these points, the cavity has a waist of about $18\,\mu$m.

	\begin{figure*} 
		\centering
		\includegraphics[width=\textwidth]{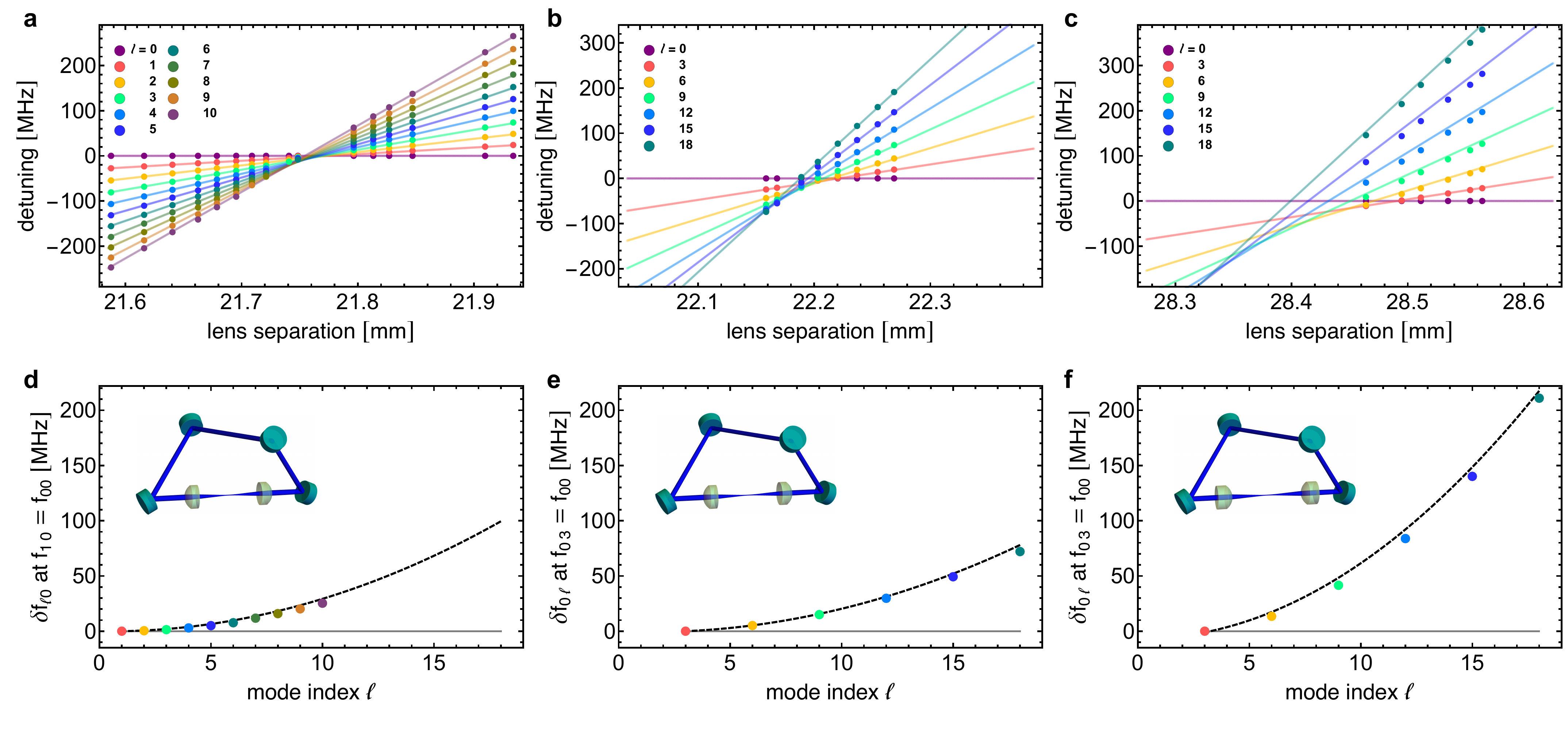}
		\caption{Measured spectra of the twisted lens cavity. (a) near the $s=1$ point, (b) near the $s=3$ point, (c) near the $s=3$ point with backwards lenses. (a-c) have the same vertical and horizontal spans. Data positions along the horizontal axis are inferred from the splitting between the fundamental mode and the lowest-excited mode in the degenerate manifold. Markers indicate measured data points, while the lines indicate the perturbative prediction with no free parameters. (d-f) Missed-degeneracy due to aberrations for the $s=1$, $s=3$ point, and $s=3$ point with backwards lenses, respectively. Mode energies relative to the fundamental mode are plotted against angular momentum $\ell$ at the lens separation where the fundamental and $\ell = s$ modes are degenerate. Without aberrations, these modes would all be degenerate at the same lens separation, so this quantity would be zero (solid line). Data is shown in colored points, obtained by linear fits to the data of (a-c). Our perturbative calculation, with no free parameters, is shown by the dashed line (shown as continuous, for ease of comparison to the data). (d-f) have the same vertical and horizontal spans. Note that the aberration contribution to the spectrum is $\propto \ell^{2}$, as expected for a quartic perturbation.
		}
		\label{fig4}
	\end{figure*}

To benchmark our calculations, we built such a cavity and measured its spectrum. Without astigmatism, we expect the degeneracy to be limited by quartic terms from spherical aberration and non-paraxial propagation. The lenses are plano-convex fused silica substrates with a 5 mm radius of curvature (ROC). They are anti-reflection coated and super-polished to $<2$ {\AA} surface roughness\footnote{Manufactured and coated by Perkins Precision Developments (PPD)}. The lenses support a cavity finesse  $\mathcal{F}\geq 18,500$ (observed). We set the finesse to 5570(10) via the in/out -coupling mirror transmissions.

We probe the spectrum of our cavity with a $\lambda = 784$~nm diode laser by overlapping two paths to inject light into the cavity. One path goes through an electro-optic modulator (EOM), and serves as a frequency ruler by exciting the $00$ mode. The other path is incident on a digital micro-mirror device (DMD). In order to measure the frequency of a higher-order mode, we can excite that mode via holographic beam shaping with the DMD ~\cite{Zupancic2016}. The EOM frequency is then tuned until the peaks from the two paths overlap, providing sub-MHz resolution (linewidth $\sim 2\pi\times 400$ kHz) of the frequency difference between the mode of interest and the $00$ mode.

Spectra at a range of lens splittings near degeneracy points can be seen in Fig.~\ref{fig4}. Fig.~\ref{fig4}a shows $s=1$, Fig.~\ref{fig4}b shows $s=3$, and Fig.~\ref{fig4}c shows $s=3$ for lenses oriented backwards\footnote{The ``lens separation" in Figs.~\ref{fig3},\ref{fig4} refers to the separation of the centers of the lens. Since the Gouy phase is set by the curved surface separation, the ``lens separation" for the backwards configuration is larger.}. This backwards configuration significantly worsens the observed aberrations, in agreement with our theory. Importantly, this effect does \emph{not} appear when modeling the lens as a position-dependent phase plate (as in Refs.~\cite{Laabs1999, visser2005spectrum, Klaassen2006}, with mirrors). The slope-dependent perturbation terms must be included to accurately reproduce the spectra. The relevant full perturbation forms can be found in Appendix~\ref{sec:ax_sym_aberrations}.

Remarkably, the $s=1$ point supports stable modes. From a purely paraxial standpoint, this cavity should be unstable: the ABCD matrix is singular, akin to an exactly-confocal cavity~\cite{siegman86}. Quadratic astigmatism resonantly couples modes in this configuration, so the cavity lenses must be aligned very precisely. A consideration using the results of Appendix~\ref{subsubsec:refraction_paraxial_astigmatism_pert} indicates that curved lens surfaces must be centered / un-tilted with respect to the cavity axis to within $\delta x\lesssim10\,\mu$m (and/or equivalent tilt $\sim \frac{\delta x}{R} \approx 0.1^{\circ} $; in practice the positioning is a more stringent constraint). Without this level of alignment, quadratic astigmatism mixes the modes near degeneracy, leading to level repulsion (as seen in, e.g., Fig.~\ref{fig1} for the cubic case). A quartic term breaks this degeneracy, confining the light to within a finite radius of the cavity axis in the presence of small but finite misalignment.

The $s=3$ point enjoys protection against quadratic astigmatism (it is off-resonant), and is thus significantly less sensitive to alignment. However, only $1/3$ of the number of modes lie within a given frequency window, as compared to $s=1$. Eventually the quartic term breaks this degeneracy, leaving only a few modes within several MHz in the ``degenerate" manifold.

Flipping the lenses such that the curved side faces the cavity waist worsens the aberrations, increasing their effect on the spectrum. This can be seen from the wider spread in zero-crossings of the modes in Fig.~\ref{fig4}c compared to Fig.~\ref{fig4}b, as well as the stronger quadratic contribution to mode energies in Fig.~\ref{fig4}f than in Fig.~\ref{fig4}e. Due to technical aspects of our alignment procedure, the backwards lens-cavity was more ambiguous to align. And while the quadratic astigmatism of misaligned lenses does not affect the stability of the $s=3$ manifold, it can affect the size of the splittings. For the model in Fig.~\ref{fig4}f, we have included a single lens tilt of $2.5^\circ$. In reality, both lenses could be tilted / displaced in an arbitrary transverse direction. This would be difficult and not so informative to disambiguate.

\begin{figure*} 
	\centering
	\includegraphics[width=\textwidth]{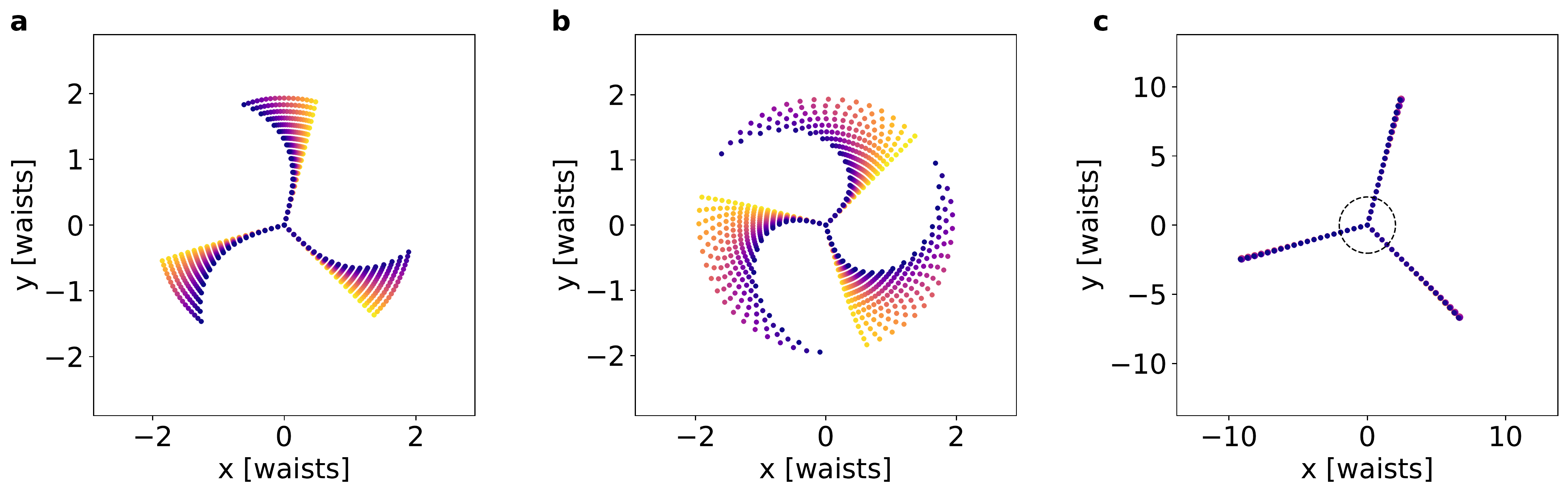}
	\caption{
		Ray tracing inside of an $s=3$ resonator. The real part of the eigenray corresponding to the degenerate index is scaled to an array of radii in the transverse plane of the cavity waist. Each initial ray is then propagated around the cavity for 50 round trips. At small radii, the ray comes back to itself after 3 round trips, indicating that the Gouy phase is $\frac{2\pi}{3}$. At increasing radii, quartic aberrations shift the mode energies, and the rays begin to precess. (a) Hit patterns for the twisted lens cavity (Fig.~\ref{fig4}b,e). (b) Hit patterns with the lenses oriented backwards(Fig.~\ref{fig4}c,f). The stronger aberration is evident in the increased change in precession rate with radius. (c) 500 round trips per radius in the same cavity as (a), but with aspheric lenses rather than spherical. The asphere surface profile is given by $z(r) = \frac{1}{2R} r^{2} + c_{4} r^{4}$, where $R=5$ mm is the lens base radius of curvature, and $c_{4} = 0.00038\text{ mm}^{-3}$ (with sign convention such that the optic has more glass than a spherical lens, and less than a parabolic lens). This aspheric profile agrees with the perturbative method prediction. The dashed inner circle indicates the maximum radius ray from (a,b). The aberration-correction is evident out to large radii.
	}
	\label{fig5}
\end{figure*}

Interestingly, we have found that numerical ray tracing is a powerful tool for identifying resonator aberration, where we take the radius-dependence of the round-trip ray precession~\cite{Klaassen2006} as an indication of uncorrected aberrations (see Fig.~\ref{fig5}). Optimizing the resonator geometry to suppress this precession has coincided with the perturbatively computed optimum in the radially-symmetric cases we have tested. This suggests that the wave-properties of these resonators arise from interference between ray round trips, an idea that merits further exploration.\\

\section{Conclusions} \label{sec:summary}
We have demonstrated a quantitative method to calculate optical resonator mode spectra beyond the paraxial, quadratic limit. These methods are especially appropriate for degenerate cavities with small waists. Degeneracy requires mode energy spacing uniformity to about 1 part in the finesse ($\sim 10^{2} - 10^{5}$). Small waist sizes, $\lesssim 50\,\mu$m for optical wavelengths, tend to require terms at higher-than-quadratic orders to achieve this level of accuracy.

We have shown that this perturbative method accurately predicts the aberrated behavior of several non-standard cavities. This includes the spectral behavior and mode profiles of a cubically-aberrated twisted cavity and quantitative spectra of a twisted cavity for various orientations of intracavity spherical lenses. This quantitative understanding will allow for improved design of degenerate cavities, particularly for use in quantum optics experiments with atoms, where a small waist size leads to stronger atom-photon interactions. For example, appropriate aspheric lenses could be employed to engineer a degenerate spectrum in the presence of aberrations.

Diffraction-limited cavity modes with high numerical aperture (NA) are an appealing target for this line of research. There is an open problem in considering yet-higher-order terms, in particular their commutators. A deeper understanding of the connection to ray-tracing may shed light on this problem. At this level, vector-optical effects may also contribute.

\onecolumngrid
\newpage
\appendix
\section{Deriving the perturbation polynomials} \label{appendix:pert_polys}
To calculate the effect of a perturbation on the resonator spectrum, there remains the question of determining the operator to exponentiate. In this Appendix, we derive these operators for some common aberrations. In doing so, the procedure is outlined for handling a general aberration, so long as its action on a ray\footnote{Following convention of the Hamiltonian optics literature, $\bm{q}$ is used to represent position (rather than $\bm{x}$), and the vertical ordering of the canonically conjugate variables within $\bm{v}$ is reversed from the ABCD matrix convention.} $\begin{pmatrix} \bm{p} \\ \bm{q} \end{pmatrix} $ is known. We begin with a brief overview of Hamiltonian optics, the approach we will use to derive the perturbation forms.

\subsection{Hamiltonian optics}

A transverse ray $\bm{v} = \begin{pmatrix} \bm{p} \\ \bm{q} \end{pmatrix}$ can be viewed as a point in phase space for transverse position $\bm{q}$ and canonically conjugate transverse momentum $\bm{p}$. The momentum is typically related to ray slope by $\bm{p} = n \bm{s}$, where $n$ is the local index of refraction. An operator $G_{f}$ transforms such a ray by

\begin{equation}
    G_{f}:\begin{pmatrix} \bm{p} \\ \bm{q} \end{pmatrix} \rightarrow \begin{pmatrix} \bm{p'} \\ \bm{q'} \end{pmatrix}
\end{equation}
\noindent
where $f = f(\bm{p},\bm{q})$ is a polynomial, and the operator $G_{f}$ is constructed by
\begin{equation} \label{eq:operator_from_poly}
    G_{f} := \exp{\left(\hat{f}(\bm{p},\bm{q})\right)}.
\end{equation}
\noindent
The hat indicates a Poisson bracket operator, such that 
\begin{equation}
    \hat{f}(\bm{p},\bm{q}) :=  \left\{ f(\bm{p},\bm{q}),\cdot \right\} = \sum_{i}
    \frac{\partial f}{\partial q_{i}} \frac{\partial}{\partial p_{i}} -
    \frac{\partial f}{\partial p_{i}} \frac{\partial}{\partial q_{i}}
\end{equation}
\noindent
Exponentiation of a hatted operator is given by:
\begin{equation} \label{eq:hatted_operator_exponentiation}
    \exp(\hat{f}) := 1 + \left\{ f,\cdot \right\}
    + \frac{1}{2!} \left\{ f,\left\{ f,\cdot \right\} \right\} + \ldots,
\end{equation}
\noindent
i.e., a formal expansion of the usual power series of exponentiation. Powers of $\hat{f}$ represent nested Poisson brackets $\{f,\circ\}^{m} = \{f,\{f,\ldots,\{f,\circ\}\}\}$ (where there are $m$ nested Poisson brackets).

Evolution is described by a transverse Hamiltonian
\begin{equation} \label{eq:transverse_Hamiltonian}
    h(\bm{p}, \bm{q}) := -\sqrt{n(\bm{q})^{2} - \bm{p}^{2}} = -p_{z}
\end{equation}
\noindent
where $\bm{p}$ and $\bm{q}$ (taken to be either 1- or 2-dimensional) follow the Hamilton equations
\begin{equation} \label{eq:Hamilton_equations}
    \frac{d q_{i}}{d z} = \frac{\partial h}{\partial p_{i}}, \qquad \qquad
    \frac{d p_{i}}{d z} = \frac{\partial h}{\partial q_{i}}.
\end{equation}

The variables $q_{i}$ and $p_{j}$ are canonically conjugate, such that $\left\{ p_{i},q_{j} \right\} = \delta_{ij}$, and $\left\{ p_{i},p_{j} \right\} = \left\{ q_{i},q_{j} \right\} = 0 $. The momentum $\bm{p}$ is related to the ray slope $\bm{s}$ and the angle $\theta$ between the ray and the optical axis by $\bm{p} = n \sin{\theta}$. The ray slope $\bm{s}$ then just expresses $\theta$ and its direction in a local coordinate system.

Quadratic optics refers to transformations on phase space generated in the above way by quadratic polynomials. For example, taking the polynomial $g_{\text{prop}} = -\frac{z}{2n}p^{2}$, the generated transformation is

\begin{align*}
    G_{g_{\text{prop}}} : \begin{pmatrix} \bm{p} \\ \bm{q} \end{pmatrix} &= e^{\hat{g}_{\text{prop}}}:\begin{pmatrix} \bm{p} \\ \bm{q} \end{pmatrix} \\
    & = \begin{pmatrix} \bm{p} \\ \bm{q} + \frac{z}{n}\bm{p} \end{pmatrix},
\end{align*}
\noindent
which shows that $g_{\text{prop}}$ represents free propagation over a distance $z$ in a medium of index $n$. Similarly, $g_{\text{lens}} = -\frac{1}{2f}q^{2}$ generates a transformation

\begin{align} \label{eq:lens_ham_op}
    G_{g_{\text{lens}}} : \begin{pmatrix} \bm{p} \\ \bm{q} \end{pmatrix} &= e^{\hat{g}_{\text{lens}}}:\begin{pmatrix} \bm{p} \\ \bm{q} \end{pmatrix} \nonumber \\
    & = \begin{pmatrix} \bm{p} - \frac{1}{f} \bm{q} \\ \bm{q}\end{pmatrix},
\end{align}
\noindent
which represents the action of an ideal thin lens with focal length $f$.

These quadratic operators generate \textit{linear} transformations on phase space, and are thus redundant with the ABCD matrix description. However unlike ray transfer matrices, the Hamiltonian optics formulation can be readily extended to \textit{non-}linear transformations of phase space by using polynomials of higher-than-quadratic order. This will be our approach to deriving the perturbation polynomials. A more complete description of these Hamiltonian optics tools can be found in Ref.~\cite{Wolf2004}. Interestingly, the operators $G_{f}$ corresponding to polynomials $f$ at a given order of the phase space variables $\bm{p}$ and $\bm{q}$ constitute a Lie group acting on that phase space. This Lie-Hamilton formulation of geometric optics is fundamental to our description and classification of aberrations.

\subsection{Non-paraxial propagation}
In taking the paraxial approximation for propagation over a distance $L$, the propagation operator 
\small
\begin{align*}
	U_{\text{prop}}(L) &\equiv \exp{\left[ i k L \sqrt{1 - \left(\frac{k_{\perp}}{k}\right)^{2}} \right]}\\
	&=  \exp{\left[ i k L \left(1 - \frac{1}{2}s^{2} - \frac{1}{8}s^{4} - \frac{1}{16}s^{6} - \frac{5}{128}s^{8} - \ldots  \right] \right)}
\end{align*}
\normalsize

\noindent
was truncated to quadratic order in ray slope $s \equiv \frac{k_{\perp}}{k}$, where $k = 2\pi / \lambda$, $k_{\perp} \equiv - i \nabla$ and $\nabla$ the gradient in the transverse plane. Free propagation doesn't couple transverse directions to each other, so we can simplify the 2D case by considering 2 copies of a 1D case.

The perturbation polynomial for non-paraxial propagation is then simply the higher-order terms of the above expansion (which can be truncated to desired order):
\begin{equation} \label{eq:f_nonparax_std}
	f_{\text{nonparax}} = -\frac{kL}{8}s^{4} - \frac{kL}{16}s^{6} - \frac{5kL}{128}s^{8} - \ldots
\end{equation}

A Hamiltonian optics consideration yields the same result. For free propagation in a medium of homogeneous index of refraction $n$, the transverse Hamiltonian (Eqn.~(\ref{eq:transverse_Hamiltonian})) is $h(p, q) = -\sqrt{n^{2}  - p^{2}} = - p_{z}$, where the $z$-axis is the optical axis. Evolution over a distance $L$ under this Hamiltonian is then generated by 
\begin{equation} \label{eq:ham_evol}
	\exp\left(- L \hat{h}(p,q)\right) : \begin{pmatrix} \bm{p} \\ \bm{q} \end{pmatrix}
\end{equation}

\noindent
using Eqn.~\ref{eq:hatted_operator_exponentiation} to evaluated the exponentiation. $\bm{p}$ and $\bm{q}$ then obey Hamilton's equations, eqs.~(\ref{eq:Hamilton_equations}):
\begin{align}
	\frac{d q_{i}}{d z} &= \frac{\partial h}{\partial p_{i}} = \frac{p_{i}}{\sqrt{n^{2} - p^{2}}} \label{eq:q_evol_free_prop} \\
	\frac{d p_{i}}{d z} &= \frac{\partial h}{\partial q_{i}} = 0  \label{eq:p_evol_free_prop}
\end{align}
\noindent
where the second equality in each case represents specialization to our case of free propagation in a homogeneous medium.

Expanding the Hamiltonian into a power series gives:
\begin{align}
	h(p, q) &= -\sqrt{n^{2}  - p^{2}} \\
			&= - n + \frac{1}{2n} p^{2} + \frac{1}{8 n^{3}} p^{4} + \frac{1}{16 n^{5}} p^{6} + \frac{5}{128 n^{7}} p^{8} + \ldots \label{eq:ham_expansion}
\end{align}

The constant term can be discarded (it does not affect the dynamics), and the quadratic term generates the paraxial (linear) transformation of phase space. Identifying the ray slope as the canonical momentum divided by the local index of refraction, $s = p/n$, collecting the remaining terms of Eqn.~(\ref{eq:ham_expansion}) and combining with Eqn.~(\ref{eq:ham_evol}) then yields:
\begin{equation}
	f_{\text{nonparax}}^{\text{h}} = -\frac{n L}{8}s^{4} - \frac{n L}{16}s^{6} - \frac{5 n L}{128}s^{8} + \ldots
\end{equation}
\noindent
This agrees with Eqn.~(\ref{eq:f_nonparax_std}) (where we neglected to include the index $n$) after reintroducing the wavenumber $k$ to reflect moving from the Hamiltonian optics' Poisson bracket structure to the operator formalism used in our computations (whose structure is determined instead by the commutation relations of the operators).

It can be verified that applying this expansion via
\begin{equation}
	\cdots : \exp{\left( -\frac{L}{16 n^{5}}p^{6} \right)} :  \exp{\left( -\frac{L}{8 n^{3}}p^{4} \right)} : \exp{\left( -\frac{L}{2 n}p^{2} \right)} : \begin{pmatrix} p \\ q \end{pmatrix}
\end{equation}

\noindent
yields the transformation
\begin{equation}
	q \rightarrow q + z p + \frac{z}{2} p^{3} + \frac{3 z}{8} p^{5} + \ldots
\end{equation}

\noindent
which is the power series of the evolution Eqn.~(\ref{eq:q_evol_free_prop}) for $q$. Thus, this polynomial indeed generates the desired transformation of phase space describing non-paraxial ray propagation.

\subsection{Spherical aberration} \label{sec:ax_sym_aberrations}
A similar procedure can be performed to find the generating polynomials for spherical aberration. A reflected or refracted ray can be solved for order-by-order following procedures outlined in Refs.~\cite{Navarro-Saad1986, Wolf2004}. The generating polynomial of these transformations can then be solved for with methods from the same references. This is an extremely tedious, but mechanical process, that we perform using \texttt{Mathematica}. For this reason, we list some of the more useful results of that process here.

We consider reflection and refraction at an axially-symmetric surface $\zeta(r) = \zeta_{2} r^{2} + \zeta_{4}r^{4} + \ldots$ . For a spherical optic with radius of curvature $R$, the coefficients $\zeta_{2n}$ are obtained by Taylor-expanding the surface profile $\zeta_{\text{sph}}(r) = R \left(1 - \sqrt{1-\left(\frac{r}{R}\right)^{2}} \right)$. Reflection and refraction transform an incident ray $\bm{v}_{\text{in}} := \begin{pmatrix} \bm{p} \\ \bm{q} \end{pmatrix}$ into an outgoing ray $\bm{v}_{\text{refl}}$ and $\bm{v}_{\text{refr}}$, respectively.

\subsubsection*{Reflection}
Up to the first beyond-paraxial order, the outgoing ray for reflection from an axially-symmetric surface $\zeta(r)$ at normal incidence is
\begin{equation} \label{eq:refl_ray_LO_trans}
	\bm{v}_{\text{refl}}  = \begin{pmatrix}
		\bm{p} - 4 \zeta_{2}\bm{q}  + 2 \zeta_{2} p^{2}\bm{q} + 4 \zeta_{2}^{2} q^{2}\bm{p} - 8 \zeta_{2}^{2} \bm{p}\cdot\bm{q}\,\bm{q} + 
		8\left( 2\zeta_{2}^{3} -\zeta_{4} \right) q^{2}\bm{q}\\
		\\
		\bm{q} - 2\zeta_{2} q^{2} \bm{p} + 4\zeta_{2}^{2} q^{2}\bm{q} 
	 \end{pmatrix} +\mathcal{O}(p^{5}, q^{5})
\end{equation}
\noindent
Recall that $\bm{p}$ and $\bm{q}$ are 2-dimensional; unbolded terms such as $q^{2} \equiv \bm{q} \cdot \bm{q}$ represent scalars.

The polynomial that generates this transformation via Eqn.~(\ref{eq:operator_from_poly}) can be written as a sum over aberration orders as $f_{\text{refl}} = f^{[4]}_{\text{refl}} + f^{[6]}_{\text{refl}} + \ldots$. The lowest two aberration order polynomials are

\begin{align}
\label{eqn:transpoly}
	f^{[4]}_{\text{refl}} &= \zeta_{2} p^{2}q^{2} - 4 \zeta_{2}^{2}\bm{p}\cdot\bm{q} \, q^{2} + 2(4\zeta_{2}^{3} - \zeta_{4})q^{4} \\
	f^{[6]}_{\text{refl}} &= 
								  \frac{\zeta_{2}}{4}  										p^{4}q^{2}
								- 2\zeta_{2}^{2}      										p^{2} \bm{p}\cdot\bm{q} q^{2}
								+ 8\zeta_{2}^{3}  											\left( \bm{p}\cdot\bm{q} \right)^{2}q^{2}
								+ 2\left( \zeta_{2}^{3} + \zeta_{4} \right)  				p^{2}q^{4}
								- 4\left( 6\zeta_{2}^{4} + \zeta_{2}\zeta_{4} \right)  		\bm{p}\cdot\bm{q}q^{4}
								+ 2\left( 12\zeta_{2}^{2}\zeta_{4} - \zeta_{6} \right) 		q^{6}
\end{align}

These formulas use the convention that a concave reflecting surface has positive radius of curvature $R$. Note that Eqn.~(\ref{eq:refl_ray_LO_trans}) is the transformation due only to $f^{[4]}_{\text{refl}}$. 

\subsubsection*{Refraction}
In contrast to the reflection formulas, in this section we take a convex refracting surface to have positive radius of curvature $R$. For refraction from index $n_{1}$ into index $n_{2}$ at an interface with surface profile $\zeta(r)$, the outgoing ray, to lowest beyond-paraxial order, is

\begin{equation} \label{eq:refr_ray_LO_trans}
	\bm{v}_{\text{refr}}  = 
	\begin{pmatrix}
		\bm{p}
			- \left( 2\left(n_{2}-n_{1}\right) \zeta_{2} \right) 															\bm{q}
			- \left( \frac{n_{2} - n_{1}}{n_{1}n_{2}} \zeta_{2}\right) 														p^{2}\bm{q}
			- \left( 2\frac{n_{2}-n_{1}}{n_{1}} \zeta_{2}^{2} \right) 														q^{2}\bm{p}
			- \left( 4\frac{n_{2}-n_{1}}{n_{2}} \zeta_{2}^{2} \right) 														\bm{p}\cdot\bm{q}\,\bm{q}
			+ \left( 4\frac{n_{2}-n_{1}}{n_{2}} \left(n_{2}\zeta_{4}-\left(n_{2}-n_{1}\right)\zeta_{2}^{3} \right) \right) 	q^{2}\bm{q}\\
		\\
		\bm{q}
			+\left( \frac{n_{2} - n_{1}}{n_{1}n_{2}}\zeta_{2} \right) q^{2} \bm{p}
			+\left( 2\frac{n_{2}-n_{1}}{n_{2}} \zeta_{2}^{2}  \right) q^{2} \bm{q} 
	\end{pmatrix} +\mathcal{O}(p^{5}, q^{5})
\end{equation}

The polynomial that generates this transformation via Eqn.~(\ref{eq:operator_from_poly}) can be written as a sum over aberration orders as $f_{\text{refr}} = f^{[4]}_{\text{refr}} + f^{[6]}_{\text{refr}} + \ldots$. The lowest two aberration order polynomials are

\begin{align}
\label{eqn:reflpoly}
	f^{[4]}_{\text{refr}} &= 
		  \left( \frac{n_{1}-n_{2}}{2n_{1}n_{2}}\zeta_{2} \right)															p^{2}q^{2}
		 +\left( 2\frac{n_{1}-n_{2}}{n_{2}}\zeta_{2}^{2} \right) 															\bm{p}\cdot\bm{q} \, q^{2}
		 +\left( \left(n_{1}-n_{2}\right) \left( 2\frac{ n_{1} - n_{2}}{n_{2}} \zeta_{2}^{3} + \zeta_{4} \right) \right) 	q^{4}
		 \\
	f^{[6]}_{\text{refr}} &= 
		\left( \frac{\left(n_1^3-n_2^3\right)}{8 n_1^3 n_2^3}\zeta _2 \right)  						p^{4}q^{2}
		+ \left( \frac{2 n_1^3-n_2 n_1^2-n_2^3}{2 n_1^2 n_2^3} \zeta_2^2\right)     					p^{2} \bm{p}\cdot\bm{q} q^{2} \nonumber\\
		&+ \left( \frac{2\left(n_1^3-n_1^2n_2 + n_1n_2^2 -n_2^3\right)}{n_1 n_2^3} \zeta_2^3 \right) 	\left( \bm{p}\cdot\bm{q} \right)^{2}q^{2}
		+ \left( \frac{\left(n_1-n_2\right) \left(2 n_1^2 \zeta _2^3 + n_2^2\zeta_4\right)}{2 n_1 n_2^3} \right) 				p^{2}q^{4} \nonumber\\
		&+ \left( \frac{2 \left(n_1-n_2\right) }{n_1 n_2^3} \left(\left(n_1-n_2\right) \left(2 n_1^2+n_1 n_2 +2 n_2^2\right)\zeta_2^3 +n_2^2 \left(2 n_1+n_2\right)\zeta_4 \right) \zeta_2 \right)  	\bm{p}\cdot\bm{q}q^{4} \nonumber\\
		&+ \left( \frac{\left(n_1-n_2\right)}{n_2^3} \left(2 \zeta_2^2 \left(n_1-n_2\right) \left(\left(n_1-n_2\right) \left(n_1+n_2\right)\zeta_2^3 +3 n_2^2 \zeta_4\right) + n_2^3 \zeta_6 \right) \right) 		q^{6}
\end{align}

 Note that Eqn.~(\ref{eq:refr_ray_LO_trans}) is the transformation due only to $f^{[4]}_{\text{refr}}$. These equations are clearly cumbersome; this is in part why numerical ray tracing is so widely used. Equations of this type can be encoded in slightly more manageable, albeit somewhat obscured, coefficient lists as in Chapters 13-14 of Ref.~\cite{Wolf2004}.

\subsection{Non-normal incidence off a curved surface} \label{sec:non_normal_aoi}

Extending the results of Section~\ref{sec:ax_sym_aberrations} beyond cases of axial symmetry significantly complicates the necessary algebra. An outline of the solution methods can be found in Ref.~\cite{GrpThrtcIII1987}, using the helicity basis of phase space, and the resulting asymmetric aberration polynomials. Despite the significant increase in complexity, this is an important case as it explains the observation of large aberrations in Ref.~\cite{clark2020observation} alongside the much smaller aberrations of Ref.~\cite{schine2016synthetic}. In both cases, the primary culprit is cubic astigmatism resulting from non-normal incidence off of curved mirrors.

For reflection at angle of incidence $\alpha$ from a spherical mirror with radius of curvature $R$ ($R>0$ is concave), the lowest-order perturbation polynomial is:
\begin{align} \label{eq:cubic_pert_poly}
	f^{[3]}_{\text{refl},\,\alpha} &= 
									 \left(\frac{n}{R^{2}}\sec^{2}\alpha \tan \alpha \right)				x^{3}
									+\left(\frac{n}{R^{2}}\cos 2\alpha \tan \alpha \right)					x y^{2} \nonumber\\
									&-\left(\frac{1}{R}\sec\alpha \tan \alpha \right)						p_{x} x^{2}
									+\left(\frac{1}{R}\sin\alpha \right)									p_{x} y^{2}
									-\left(\frac{2}{R}\sin\alpha \right)									p_{y} x y
\end{align}

As with the spherical aberration polynomials, an enormous amount of algebra is required to get from the recipe of Ref.~\cite{GrpThrtcIII1987} to Eqn.~(\ref{eq:cubic_pert_poly}). We use \texttt{Mathematica} for this purpose. To our knowledge, this is the first presentation of this result.

\subsection{Paraxial astigmatism}
It is well-known that a confocal cavity becomes unstable when astigmatism is introduced. It is useful to explain this in several pictures. In the ray transfer matrix picture, this can be seen from the eigenvalues of the round-trip ABCD matrix having norms not equal to 1. In the operator picture, we have the relation $\hat{\nu} = B \hat{\xi}$, where $\hat{\nu} = \begin{pmatrix} x & y & s_{x} & s_{y} \end{pmatrix}^{\intercal}$ and $\hat{\xi} = \begin{pmatrix} a & b & a^{\dagger} & b^{\dagger} \end{pmatrix}^{\intercal}$. Paraxial astigmatism, represented by the polynomial (or operator) $x^{2} - y^{2}$ can then be seen to contain such terms as:
\begin{equation}
	x^{2} - y^{2} \ni aa, \, bb, \, ab, \, a^{\dagger}a^{\dagger}, \, b^{\dagger}b^{\dagger}, \, a^{\dagger}b^{\dagger}.
\end{equation}

These terms connect levels differing by two excitations. For the confocal resonator, these modes have the same energy, and thus the coupling is resonant. This mode-mixing leads to instability, as higher- and higher-order modes mix with lower-order modes. Finally, in a ray-tracing picture, this can be seen by the hit patterns of subsequent cavity round trips drifting towards infinity along equipotentials of the astigmatic $x^{2} - y^{2}$ potential.

However, when considering higher order terms, this can actually be stabilized against. For example, in a resonator with paraxial astigmatism \emph{and} quartic spherical aberration, the quartic term grows quadratically with mode index, while the paraxial term grows only linearly. Thus, the degeneracy leading to the resonant mode-mixing will eventually be cut off by the quartic term.

To describe such a case, it will be useful to treat the paraxial astigmatism as a perturbation, even though it can sometimes be handled within the ABCD matrix formalism. To do so, we will write the operation as a stigmatic ABCD matrix, and a perturbation polynomial capturing the astigmatism.

\subsubsection{Reflection}
For reflection from an astigmatic curved surface with sagittal and tangential radii of curvature $R_{y}$ and $R_{x}$, respectively, the ABCD matrix is~\cite{Massey1969, siegman86}:
\begin{equation} \label{eq:M4_Rx_Ry}
M = \begin{pmatrix}
1 & 0 & 0 & 0 \\
0 & 1 & 0 & 0 \\
-\frac{2}{R_{x}} & 0 & 1 & 0 \\
0 & -\frac{2}{R_{y}} & 0 & 1 \\
\end{pmatrix}
\end{equation}
\noindent
where $M$ acts on the vector $\begin{pmatrix}  x & y & p_{x} & p_{y}\end{pmatrix}^{\intercal}$. This transformation is generated by exponentiation under the Poisson bracket by the polynomial $-\frac{1}{R_{x}} x^{2} - \frac{1}{R_{y}} y^{2}$. If we take $R_{x} = R$ and $R_{y} = R + \delta R$, this we can simply separate this generating polynomial into a stigmatic, paraxial curved mirror with radii of curvature $R_{x} = R_{y}=R$, plus a perturbation given by

\begin{equation}
f^{[2]}_{\text{refl}} =  \frac{\delta R}{R \left( R + \delta R \right)} y^{2}
\end{equation}

This perturbation could equivalently be expanded into a term $\propto \left(x^{2} + y^{2}\right)$ (which slightly modifies the transverse trapping of the mirror) and term $\propto \left(x^{2} - y^{2}\right)$, which is the manifestly astigmatic term. A common example of this situation is incidence on a spherical reflector (with radius of curvature $R$) at an angle $\theta$. In this case, the sagittal radius of curvature ($R_{S} = R/\cos\theta$) and tangential radius of curvature ($R_{T} = R\cos\theta$) are used in Eqn.~(\ref{eq:M4_Rx_Ry}) to describe the paraxial action of the optic~\cite{siegman86}.

\subsubsection{Refraction} \label{subsubsec:refraction_paraxial_astigmatism_pert}
For astigmatic refraction, the process is similar, but with a more complicated paraxial transformation, which is derived in Ref.~\cite{Massey1969}. For refraction at a curved interface with sagittal and tangential radii of curvature $R_{S}$ and $R_{T}$, respectively, at an angle of incidence $\theta$, the ABCD matrix is
\begin{equation} \label{eq:paraxial_tilted_refraction}
M = \begin{pmatrix}
\frac{\sqrt{n_{r}^{2}+\cos^{2}\mkern-4mu\theta-1}}{n_{r} \cos\theta} & 0 & 0 & 0 \\
0 & 1 & 0 & 0 \\
\frac{n_{r}\left(\cos\theta - \sqrt{n_{r}^{2}+\cos^{2}\mkern-4mu\theta-1}\right)}{R_{T}\cos\theta \sqrt{n_{r}^{2}+\cos^{2}\mkern-4mu\theta-1}} & 0 & \frac{n_{r} \cos\theta}{\sqrt{n_{r}^{2}+\cos^{2}\mkern-4mu\theta-1}} & 0 \\
0 & \frac{\cos\theta - \sqrt{n_{r}^{2}+\cos^{2}\mkern-4mu\theta-1}}{R_{S}} & 0 & 1 \\
\end{pmatrix}
\end{equation}
\noindent
In this expression, $n_{r} = \frac{n_{2}}{n_{1}}$ for refraction from a medium with index $n_{1}$ into a medium of index $n_{2}$. This matrix acts on the vector $\begin{pmatrix}  x & y & p_{x} &  p_{y} \end{pmatrix}^{\intercal}$, where $y$ indicates the sagittal axis and $x$ indicates the tangential axis. 

The sagittal and tangential directions can be treated as independent 1D problems, as the above matrix does not couple them. For each axis, we can perform an Iwasawa decomposition (see Ref.~\cite{Simon1998} and Chapter 9.5 of Ref.~\cite{Wolf2004}). In general, this decomposes a linear transformation of phase space $M \in \mathsf{Sp}(2, \mathbb{R})$ as

\begin{equation}
    M = K \,\, A \,\, N
\end{equation}

\noindent
where $K = \begin{pmatrix} \cos\omega & -\sin\omega\\ \sin\omega& \cos\omega\end{pmatrix}$ is a fractional Fourier transform with angle $\omega$, $A = \begin{pmatrix} e^{\alpha} & 0\\ 0& e^{-\alpha}\end{pmatrix}$ is a pure magnifier with magnification $e^{-\alpha}$, and $N = \begin{pmatrix} 1 & G\\ 0& 1\end{pmatrix}$ is a thin lens with strength $G$. Following the convention of Hamiltonian optics, the matrices of this paragraph, as written, act on the vector $\begin{pmatrix}  p \\ q \end{pmatrix}$.

The sagittal component for the refraction of Eqn.~(\ref{eq:paraxial_tilted_refraction}) is simply a thin lens of strength $G_\text{sagittal} = \frac{\cos\theta - \sqrt{n_{r}^{2}+\cos^{2}\mkern-4mu\theta - 1}}{R_{S}} \approx \frac{1-n_{r}}{R_{S}} + \frac{ \left( 1 - n_{r}\right)}{2 n_{r} R_{S}}\theta^{2}$, to lowest order in $\theta$. That is: $\omega = 0$, $\alpha = 0$, and the ABCD matrix for the sagittal axis is just given by 
\begin{equation}
M_{\text{sagittal}} = \begin{pmatrix}
1 & G_\text{sagittal} \\
0 & 1 \\
\end{pmatrix}
\end{equation}

The polynomial that generates this transformation via exponentiation of the Poisson bracket (see Eqn.~(\ref{eq:lens_ham_op})) is
\begin{equation}
f_{\text{sagittal}} = \frac{1}{2} G_\text{sagittal} \, y^{2} .
\end{equation}

The tangential axis is a bit more complicated, and must be written as a composition of a thin lens and a magnifier (i.e., only $\omega=0$). The tangential ABCD matrix is decomposed as

\begin{equation}
M_{\text{tangential}} = \begin{pmatrix}
e^{\alpha_{\text{tangential}}} & 0 \\
0 & e^{-\alpha_{\text{tangential}}} \\
\end{pmatrix}
\begin{pmatrix}
1 & G_\text{tangential} \\
0 & 1 \\
\end{pmatrix}
\end{equation}
\noindent
with $G_\text{tangential} = \frac{1-n_{r}}{R} + \frac{ \left( 1 - n_{r}^{3} \right)}{2 n_{r}^{2} R}\theta^{2}$ and $e^{\alpha_{\text{tangential}}} = \frac{1}{2}\left( \frac{1}{n_{r}^{2}} -1 \right) \theta^{2}$, both to lowest order in $\theta$. The polynomial that generates this transformation under exponentiation of the Poisson operator is then seen to be:

\begin{equation}
f_{\text{tangential}} = \frac{1}{2} G_\text{tangential} \, x^{2}  + \alpha_{\text{tangential}} \, p_{x} x
\end{equation}
\noindent
because $e^{\alpha\,p\,q} \begin{pmatrix}  p \\ q \end{pmatrix} = \begin{pmatrix} e^{\alpha} \,\,\,\,\,  p \\ e^{-\alpha} \, q \end{pmatrix}$.

Finally, we can split the total generating polynomial into a stigmatic paraxial transformation, and the astigmatic perturbation. For a spherical lens ($R_{S} = R_{T} = R$), we get

\begin{equation}
f_{\text{tot}} = f_{\text{sagittal}} + f_{\text{tangential}} = f_{\text{paraxial}} + f^{[2]}_{\text{refrac}}
\end{equation}
\noindent
where $f_{\text{paraxial}} = \frac{1-n_{r}}{2 R} \left( x^{2} + y^{2} \right)$ is the stigmatic paraxial transformation that generates the ABCD matrix 

\begin{equation}
M = \begin{pmatrix}
1 & 0 & 0 & 0 \\
0 & 1 & 0 & 0 \\
\frac{1-n_{r}}{2 R} & 0 & 1 & 0 \\
0 & \frac{1-n_{r}}{2 R} & 0 & 1 \\
\end{pmatrix}
\end{equation}
\noindent
and the perturbation polynomial is 

\begin{equation}
f^{[2]}_{\text{refrac}} = \frac{ \left( 1 - n_{r}^{3} \right) \theta^{2} }{ 4 n_{r}^{2} R} x^{2} 
+ \frac{ \left( 1 - n_{r}     \right) \theta^{2} }{ 4 n_{r} R    } y^{2} 
+ \frac{ \left( 1 - n_{r}^{2}     \right) \theta^{2} }{ 2 n_{r}^{2}  } p_{x} x.
\end{equation}

Note that this expression can be used for off-center lenses as well, by considering the incident beam to simply be at an angle to the normal vector at the point of contact with the front and back surfaces of the lens.

\section{Closed-Form Expression for Symmetric Aberrations in a Symmetric Twisted Resonator} \label{appendix:closed_form_twisted}

For a resonator whose paraxial, quadratic approximation exhibits a degeneracy or near-degeneracy in only one of its two transverse quantum numbers, and only for every third mode~\cite{schine2016synthetic}, the role of quartic aberrations takes a particularly simple form. This is because quartic aberrations, generically, consist of all or nearly all possible combinations of four raising/lowering operators, and as such, can mix every fourth mode along one axis, increment/decrement one mode index by one or two, or increment/decrement one mode index by one or two while decrementing/incrementing the other by three or two, etc... If the near-degenerate mode manifold only consists of states with fixed index for one of the two quantum numbers, the aberration cannot change that index (doing so would not conserve energy) so most of the aberration terms are thus disallowed. There remain terms that increase, and then decrease, the index that must remain fixed, but these terms are then \emph{quadratic} in the other index (the one for the degenerate manifold), and amount to a renormalization of the trapping that can be tuned away by slightly adjusting the resonator parameters (typically its length). We are finally left with: terms that (in net) increase or decrease the degenerate index by four, or two, both of which are disallowed by the degeneracy of the manifold (remember: only every third mode is degenerate!); and finally terms which increase/decrease and then decrease/increase the degenerate index by two. This last family of terms is allowed, and results not in mode mixing, but in a mode-dependent energy shift which is \emph{quadratic} in the mode index. These are the aberrations that we were searching for.

At last, we will show that the strength of this aberration is related to the imaginary-part of the resonator $q$ parameter defined in the plane of the perturbation for non-paraxial corrections, and the imaginary part of $1/q$ for spatial perturbations. Because the only allowed term is a quadratic shift, we suggest that a simple quartic correction plate can compensate for the aberrations of a twisted cavity.

\subsection{The Calculation}
To begin, we note that for perturbations $D_k=e^{i\psi_k}$ which are ``small'', we can approximate $D_k\approx 1+i\psi_k$, then:

\begin{equation} \label{eqn:aberrationEQN}
U_{\text{rt}}=Q_{\text{rt}}\times\prod_{k=0}^N D_k\approx Q_{\text{rt}}\times \left(1+i\sum_{k=0}^{N}\psi_k\right)
\end{equation}

That is, so long as the perturbation per round-trip is small compared to $\pi$, the perturbations on the perturbations can be ignored.

From here, we consider a resonator with no astigmatism anywhere in the path, and a round-trip twist of angle $\theta$. This is likely either achieved by employing curved mirrors whose off-axis incidence is compensated by \emph{actual} astigmatism of the mirror form (technically very challenging, due to the need to superpolish an astigmatic concave form), or all-planar mirrors, and intra-cavity lenses (more practical). In either case, the round-trip ABCD matrix for either axis, excluding the twist, is a $2\times 2$ matrix $M_2$; the full $4\times 4$ round-trip ABCD matrix, including twist, is given by:

\begin{equation*}
M_4=\left(
\begin{array}{c|c}
M_2 & 0\\
\hline
0 & M_2
\end{array}
\right)\times \left[\mathbb{1} \otimes\left(\begin{array}{c c}\cos\theta & \sin\theta\\-\sin\theta & \cos\theta\end{array}\right)\right]
\end{equation*}

Because the ABCD matrix only mixes the $\hat{x}$ and $\hat{y}$ axes through a rotation, it is possible and natural to change to the chiral decoupled basis. $r_{\pm}\equiv \frac{1}{\sqrt{2}}(x\pm i y)$, $s_{\pm}\equiv \frac{1}{\sqrt{2}}(s_x\pm i s_y)$. The $2\times 2$ round-trip ABCD matrices for $\left(\begin{array}{c} r_{\pm}\\s_{\pm}\end{array}\right)$ are given by $M_{\pm}\equiv e^{\pm i\theta} M_2$. Note that one needs to be a bit careful, as, counter-intuitively, $[r_\pm,s_\pm]=0$, but $[r_\pm,s_\mp]=i\lambdabar$.

We next consider perturbations in the $k^{\text{th}}$ plane of the form $\psi_k=\alpha^0_k r^4+\alpha^1_k s^2 r^2+\alpha^2_k (s\cdot r) r^2 +\beta_k s^4$: the first three terms describe spherical aberration (see Eqn.~\ref{eqn:transpoly} and Eqn.~\ref{eqn:reflpoly}), while the final term quantifies non-paraxial propagation (for $\beta=-\frac{k l}{8}$ for propagation over a distance $l$). A simple calculation then reveals that $r^2=2r_+r_-$, $s^2=2s_+s_-$, and $s\cdot r=s_+ r_-+s_-r_+$.

To make further progress, we will write $r^4$, $s^4$, $s^2 r^2$ and $(s\cdot r)r^2$ in terms of the raising and lowering eigen-operators of the $k^{\text{th}}$ plane. The key realization is that when the aberrations in the $k^{\text{th}}$ plane are written in terms of the raising/lowering operators in that plane, Eqns.~\ref{eqn:definingplaneraisinglowering}~and~\ref{eqn:fullaberrationresult} indicate that to write this aberration in the $0^{\text{th}}$ reference plane we simply replace the raising/lowering operators with their counterparts in the reference plane: $a_k^\dagger/a_k\rightarrow a^\dagger/a, b_k^\dagger/b_k\rightarrow b^\dagger/b$.

Because $M_4$ is diagonal in the circular basis, we write the raising/lowering operators in this basis: 
\begin{eqnarray}
{u_\pm^{k}}^{\dagger}=\frac{1}{\sqrt{2}}(r_\pm/r_0^k\pm i s_\pm/s_0^k)\\
u_\pm^k        =\frac{1}{\sqrt{2}}(r_\mp/{r_0^k}^*\mp i s_\mp/{s_0^k}^*)\nonumber
\end{eqnarray}
\noindent
with $r_0^k/s_0^k$ constants that depend upon the round-trip ABCD matrix referenced to the $k^{\text{th}}$ plane as described below, for now taken as fixed but potentially complex. We can invert these relationships to write:

\begin{eqnarray}
r_{\pm}=\frac{1}{\sqrt{2} }(\tilde{r}_0^k {u_\pm^k}^\dagger+\tilde{r}_0^k{}^{*}u_\mp^k)\\
s_{\pm}=\frac{1}{\sqrt{2}i}(\tilde{s}_0^k {u_\pm^k}^\dagger-\tilde{s}_0^k{}^{*}u_\mp^k)\nonumber
\end{eqnarray}
\noindent
where $\tilde{r}_0^k\equiv 2 r_0^k \left(1+\frac{r_0^k s_0^k{}^*}{r_0^k{}^* s_0^k}\right)^{-1}$, $\tilde{s}_0^k\equiv 2s_0^k\left(1+\frac{r_0^k{}^* s_0^k}{r_0^k s_0^k{}^*}\right)^{-1}$, which reduce to $\tilde{r}_0^k=r_0^k$, $\tilde{s}_0^k=s_0^k$ when both $r_0$ and $s_0$ are real (which is \emph{only} the case in the focal/waist planes of the cavity).

We now assume (without loss of generality) that $u_-^k{}^\dagger/u_-^k$ move the mode index between Landau levels, and $u_+^k{}^\dagger/u_+^k$ move between cones/within the Landau level, so that $(u_+^k{}^\dagger)^3,(u_+^k)^3$ move within the same cone/Landau level (for an $s=3$ cone). Accordingly, an operator $\hat{P}$ that projects into a fixed Landau-level and cone, when applied to $r^4$, yields (keeping the energy-non-conserving terms produces higher-order corrections which we ignore along with higher order terms in the expansion of $e^{i\psi_k}$):
\begin{align}
\hat{P}r^4\hat{P}=\frac{|\tilde{r}_0|^4}{4}(u_+^k{}^\dagger)^2(u_+^k)^2 + \mathcal{O}\left(u_+^k{}^\dagger u_+^k\right) + \text{const}.
\end{align}

The term proportional to $u_+^k{}^\dagger u_+^k$ renormalizes the trapping and can be compensated with resonator parameters, and the constant offset has no effect at all. We now drop both of these, and define $n_\pm^k\equiv u_\pm^k{}^\dagger u_\pm^k$. Following similar calculations for $\hat{P}s^4\hat{P}$, $\hat{P}s^2 r^2\hat{P}$ and $\hat{P}(s\cdot r) r^2\hat{P}$ we are left with (after noting that $n_\pm^k\rightarrow n_\pm$ after paraxial propagation to a common reference plane):
\begin{equation}
    U_{\text{rt}}\approx Q_{\text{rt}}\times \exp\left[i\sum_{k=0}^{N}\left(\alpha^0_k |\tilde{r}_0^k|^4+\alpha^1_k |\tilde{r}_0^k|^2|\tilde{s}_0^k|^2+\alpha^2_k|\tilde{r}_0^k|^2\textrm{Im}{(\tilde{s}_0^k\tilde{r}_0^{k*})}+\beta_k|\tilde{s}_0^k|^4\right)n_+(n_+-1)\right]
\end{equation}

Next, we need to compute $\tilde{r}_0^k$/$\tilde{s}_0^k$, the zero-point motion/slope of the light in the $k^{\text{th}}$ plane. Note that we have already done this, albeit in a more complicated and general context, in Eqn.~\ref{eqn:raisingoperators}. We now consider explicitly the simpler case of raising/lowering eigen-operators of the evolution corresponding to a 2D round-trip ABCD matrix $M$: Defining the eigen-operator (with normalization $N$): $a^\dagger\equiv N\begin{pmatrix}\epsilon && \delta \end{pmatrix}\begin{pmatrix} x\\ s\end{pmatrix}$, under a round-trip $a^\dagger$ becomes $N\begin{pmatrix}\epsilon && \delta \end{pmatrix}M\begin{pmatrix} x\\ s\end{pmatrix}$. As such, $\begin{pmatrix}\epsilon && \delta \end{pmatrix}$ must be a \emph{left} eigenvector of M (or equivalently a right eigenvector of $M^\intercal$). Assuming that it is, we have $a^\dagger=N\epsilon x+N\delta s$. The normalization condition that $[a,a^\dagger]=1$ implies (using $[x,s]=i\lambdabar$, and assuming $N$ real) $i N^2\lambdabar(\epsilon^{*}\delta-\delta^*\epsilon)=1$, or $N=\sqrt{\frac{\pi}{\lambda \textrm{Im}\left[\delta^{*}\epsilon\right]}}$. This leaves us with the normalized operator:
$a^\dagger=\sqrt{\frac{\pi}{\lambda\,\textrm{Im}[\delta^{*}\epsilon]}}(\epsilon x+\delta s)$.

We can now identify:

\begin{eqnarray}
    r_0=\frac{\sqrt{\lambdabar\,\textrm{Im}[\delta^*\epsilon]}}{\epsilon}\\ s_0=\frac{\sqrt{\lambdabar\,\textrm{Im}[\delta^*\epsilon]}}{i\delta}\nonumber
\end{eqnarray}

It is informative to relate $r_0$ and $s_0$ in a particular plane to the beam $q$ parameter in that plane. We find the $q$ parameter by noting that it defines the lowest eigenmode of the cavity, which is thus annihilated by the lowering operator $a$. That is: $a e^{-i\frac{k}{2q}x^2}=0$, which yields $q=-\delta^*/\epsilon^*$. A bit of algebra then reveals that $|r_0|^2=\lambdabar|\textrm{Im}[q]|$, $|s_0|^2=\lambdabar|\textrm{Im}[q^{-1}]|$, and then that $|\tilde{r}_0|^2=\lambdabar|\textrm{Im}[q^{-1}]|^{-1}$, $|\tilde{s}_0|^2=\lambdabar|\textrm{Im}[q]|^{-1}$, $\textrm{Im}[\tilde{s}_0\tilde{r}_0^*]=\lambdabar\frac{\textrm{Re}[q]}{\textrm{Im}[q]}$.

Our final expression for the round-trip operator is thus (with $q_k$ the q-parameter of the lowest paraxial/quadratic mode in the $k^{\text{th}}$ plane of the resonator):\\

\begin{align}
    \label{eqn:qform}
    \boxed{
    U_{\text{rt}}\approx Q_{\text{rt}}\times \exp\left[{i\lambdabar^2}\sum_{k=0}^{N}\left(\alpha^0_k |\textrm{Im}[q_k^{-1}]|^{-2}+\alpha^1_k |\textrm{Im}[q_k^{-1}]\textrm{Im}[q_k]|^{-1}+\alpha^2_k |\textrm{Im}[q_k^{-1}]\textrm{Im}[q_k]|^{-1}\textrm{Re}[q_k]+\beta_k|\textrm{Im}[q_k]|^{-2}\right)n_+(n_+-1)\right]
    }
\end{align}\\

Rather than transforming the operators from the reference plane to the plane of the perturbation, it is possible to arrive at an equivalent result by instead transforming all perturbations to the reference plane, and then writing the operators in terms of the raising/lowering operators in that plane. This is achieved by making the replacement $\left(\begin{array}{c}x\\s\end{array}\right)\rightarrow M_k \left(\begin{array}{c}x\\s\end{array}\right)$ in $\psi_k$ of Eqn.~\ref{eqn:aberrationEQN}, where $M_k\equiv\left(\begin{array}{c c}A_k & B_k\\C_k & D_k\end{array}\right)$ is the ABCD matrix that propagates from the reference plane to the $k^{\text{th}}$ plane. If we further assume that the reference plane is in fact a mode \emph{waist}, then $r_\pm=\frac{r_0}{\sqrt{2}}(u_\pm^\dagger+u_\mp)$, $s_\pm=\frac{s_0}{i\sqrt{2}}(u_\pm^\dagger-u_\mp)$, with $\frac{r_0}{\sqrt{2}}=w_0$ the (real) resonator $1/e^2$ intensity radius (``waist'') of the lowest transverse mode, and $s_0=\frac{\sqrt{2}\lambdabar}{w_0}$.

Now the $r^4=4(r_+r_-)^2$ perturbation becomes $4\left[(A_k r_++B_k s_+)(A_k r_-+B_k s_-)\right]^2\equiv \chi_{0}$. We can then write:

\begin{equation}
    \hat{P}\chi_{0}\hat{P}={n_+(n_+-1)}\left[A_k^2 w_0^2+B_k^2\frac{4\lambdabar^2}{w_0^2}\right]^2+\mathcal{O}(n_+)+\text{const}
\end{equation}

For $s^4$, which becomes $4\left[(C_k r_++D_k s_+)(C_k r_-+D_k s_-)\right]^2\equiv \chi_\beta$, we find:

\begin{equation}
    \hat{P}\chi_\beta\hat{P}={n_+(n_+-1)}\left[C_k^2 w_0^2+D_k^2\frac{4\lambdabar^2}{w_0^2}\right]^2+\mathcal{O}(n_+)+\text{const}
\end{equation}

Repeating this procedure for $s^2 r^2$ ($\equiv\chi_{1}$) and $(s\cdot r) r^2$ ($\equiv\chi_{2}$) yields:

\begin{equation}
    \hat{P}\chi_{1}\hat{P}={n_+(n_+-1)}\left[A_k^2 w_0^2+B_k^2\frac{4\lambdabar^2}{w_0^2}\right]\left[C_k^2 w_0^2+D_k^2\frac{4\lambdabar^2}{w_0^2}\right]+\mathcal{O}(n_+)+\text{const}
\end{equation}
\begin{equation}
    \hat{P}\chi_{2}\hat{P}={n_+(n_+-1)}\left[A_k^2 w_0^2+B_k^2\frac{4\lambdabar^2}{w_0^2}\right]\left[A_k C_k w_0^2+B_k D_k\frac{4\lambdabar^2}{w_0^2}\right]+\mathcal{O}(n_+)+\text{const}
\end{equation}

Dropping terms which are independent-of- or linear-in- $n_+$ and plugging into $U_{\text{rt}}$ yields:

\begin{align}
\label{eqn:fullformabberation}
	U_{\text{rt}}\approx \exp{i\left[\chi_+n_++\chi_-n_--n_+(n_+-1)\sum\limits_{k=0}^{N}\left(\alpha^0_k\left[A_k^2 w_0^2+B_k^2\frac{4\lambdabar^2}{w_0^2}\right]^2+\alpha^1_k\left[A_k^2 w_0^2+B_k^2\frac{4\lambdabar^2}{w_0^2}\right]\left[C_k^2 w_0^2+D_k^2\frac{4\lambdabar^2}{w_0^2}\right]\right.\right.}\nonumber\\
	\left.\left.
	+\alpha^2_k\left[A_k^2 w_0^2+B_k^2\frac{4\lambdabar^2}{w_0^2}\right]\left[A_k C_k w_0^2+B_k D_k\frac{4\lambdabar^2}{w_0^2}\right]+\beta_k\left[C_k^2 w_0^2+D_k^2\frac{4\lambdabar^2}{w_0^2}\right]^2\right)\right]
\end{align}

The two results Eqn.~\ref{eqn:qform} and Eqn.~\ref{eqn:fullformabberation} are equivalent, though the former is more compact while the latter is more explicit.


\section*{Acknowledgements}
This work was supported by AFOSR Grant FA9550-18-1-0317, and AFOSR MURI FA9550-19-1-0399. We would like to thank Bernardo Wolf, Jerome Degallaix, and Daewook Kim for stimulating discussions.

\bibliographystyle{naturemag}
\bibliography{TheBibliography}
\end{document}